\documentclass[reprint,amsmath,amssymb,aps]{revtex4-1}
\usepackage{graphicx}
\usepackage{dcolumn}
\usepackage{bm}
\usepackage{color}
\usepackage{xcolor}

\usepackage[mathlines]{lineno}
\newcommand{\Me}{$\mathrm{Me}$}

\begin{document}
\title{Statistical self-organization of \textcolor{black}{a gas} of interacting walking drops \textcolor{black}{in a confining potential}}

\author{Adrien H\'elias}
\affiliation{Gulliver, UMR CNRS 7083, ESPCI Paris, Universit\'e PSL, 75005 Paris, France}
\affiliation{Current address: Department of Physics and Astronomy, Western University, 1151 Richmond St, N6A 3K7 London, Canada}

\author{Matthieu Labousse}
\affiliation{Gulliver, UMR CNRS 7083, ESPCI Paris, Universit\'e PSL, 75005 Paris, France}

\begin{abstract}
A drop bouncing on a vertically-vibrated surface may self-propel forward by \textcolor{black}{standing} waves and travels along a fluid interface. This system called {\it walking drop} forms a non-quantum wave-particle association at \textcolor{black}{the} macroscopic scale. The dynamics of one particle has triggered many investigations and has resulted in spectacular experimental results in the last decade. We investigate numerically the dynamics of \textcolor{black}{a gas} of walkers, i.e. a large number of walking drops evolving on a unbounded fluid interface in \textcolor{black}{the} presence of a confining potential acting on the particles. We show that even if the individual trajectories are erratic, the system presents well-defined ordered internal structure that remains invariant to \textcolor{black}{parameter variations such as the number of drops, the memory time and the bath radius}. We rationalize such non-stationary self-organization \textcolor{black}{in terms of} the symmetry of the waves and show that oscillatory pair potentials form a wavy collective state of active matter.
\end{abstract}

\maketitle

\section{Introduction}
In a seminal article, Y. Couder and E. Fort~\cite{Couder2005} have shown that a submillimetric drop may bounce on a vertically-vibrated fluid interface. Above a certain fluid acceleration threshold the drop may self-propel and be guided by standing waves~\cite{Bush2015,Bush2021,JWalker,JFM_Suzie,from_bouncing_to_floating,Vandewalle_PRL,Eddi2011,Molacek2013a,Molacek2013b,Oza_JFM_1_2013,Milewski2015,Durey2017} which are the footprint of Faraday waves~\cite{Faraday1831,Miles_1990,Benjamin1954}. The system has triggered a flurry of thought-provoking experiments mimicking effects which were previously thought to be peculiar to the quantum scale~\cite{Bush2015,Bush2021}. Specifically, the dynamics of one drop has been investigated in many configurations, like moving through a slit and double slits~\cite{Couder_Diffraction,Pucci2018,EllegaardLevinsen2020}, in cavities~\cite{Harris2013,Gilet2014,Gilet2016,Saenz2018,durey_milewski_wang_2020}, with a Coriolis force~\cite{Fort2010,Labousse2016a,Oza2014}, in a harmonic potential~\cite{Perrard2014a,Perrard2014,Labousse2014,Durey2017,Labousse2016a}, exhibiting Friedel-like oscillations~\cite{Saenz2020_friedel}, enabling statistical projections~\cite{Saenz2018}. In the last decade, \textcolor{black}{an important effort of research has led to a better understanding of both the emergence of wave-like statistics and the mechanisms responsible of classical quantizations at a macroscopic scale~\cite{Bush2015,Bush2021}}.\\
\\
In a parallel stream of research, it has been shown\textcolor{black}{, numerically and experimentally,} that complex dynamical behaviors such as memory-induced diffusion,
and run and tumbling dynamics~\cite{Hubert2019,Bacot2019,Durey2018,Durey2020,Durey_chaos,Durey_oscillation,Devauchelle,Rahil,Hubert2021} appear for a single droplet system, in which the degrees of freedom in the wave field play the role of a tailored thermal bath~\cite{Hubert2021}. \textcolor{black}{Generalized pilot-wave models~\cite{turton2018review} have shown that these behaviours are observed in parameter regimes beyond those accessible in the laboratory~\cite{Durey_chaos}.}\\
\\
The dynamics of two drops has also been investigated and promenade modes~\cite{Borghesi2014,Durey2017,JFM_Suzie,Valani2018,Arbelaiz2018,Couchman2019}, as well as quantized orbiting states have been observed and rationalized~\cite{Couder2005,Eddi2012,Durey2017,Oza2019_orbits,Couchman2019}. \textcolor{black}{In addition, correlated motions of two walkers trapped in a coupled set of cavities has led to analogs of superradiance~\cite{Papatryfonos2022,Papatryfonos2023} and the violation of static Bell's inequality~\cite{Bell_Papatryfonos}}. As for the many drops' dynamics, crystalline structures have been obtained in 2D~\cite{Eddi2009a,Eddi2011b,CouchmanEvansBush2022}, and the collective dynamics in toroidal channel ~\cite{Filoux2015,Filoux2017} as well as the spatially periodic potentials~\cite{Vandewalle_bragg} have been investigated in the linear and non-linear regime~\cite{Couchman2020,Thomson2020,Thomson2020a,Thomson2021,Barnes2020}. Recently, \textcolor{black}{drops trapped in circular cavities distributed over a two-dimensional lattice} have shown the spontaneous emergence of long distance motion synchronisation in analogy with ferromagnetism and antiferromagnetism~\cite{Saenz2021}. \textcolor{black}{In all these investigations~\cite{Eddi2009a,Eddi2011b,CouchmanEvansBush2022,Filoux2015,Filoux2017,Couchman2020,Thomson2020,Thomson2020a,Thomson2021,Barnes2020,Saenz2021}},  and because of the complexity of preventing the coalescence during drop collisions, the drops are carefully \textcolor{black}{initially} placed in either particular positions or specific relative distances to each other. \textcolor{black}{As a result of these constraints, and although it would be a remarkable playground for active matter and statistical physics, a {\it gas of walkers}, i.e. many interacting walking drops confined in a single large domain (see Fig.~\ref{Fig:fig1})} is experimentally unexplored. \textcolor{black}{A numerical investigation is a necessary first step to trigger and motivate further experimental realisations.} In this article, we numerically investigate the dynamics of a large number of walking drops evolving on a unbounded fluid interface in the presence of a confining \textcolor{black}{wall} potential acting on the particles.\\
\\
The article is organised as follows. In Sec.~\ref{Sec:II}, we introduce the numerical model. Then, in Sec.~\ref{Sec:III}, we characterize and rationalize the properties of the dynamical phases and propose a mechanism to predict the main properties of this statistical quantized self-organization. Finally, we conclude in Sec.~\ref{Sec:IV}.

\section{Numerical model and methods\label{Sec:II}}
We perform a numerical simulation of many walking drops in two dimensions. Several models~\cite{Fort2010,Oza_JFM_1_2013,Milewski2015,Rios19,turton2018review}, varying in their strategies to solve \textcolor{black}{the fluid equation, the impact conditions, and the horizontal dynamics of the drops}, have been proposed. They are in fairly good qualitative agreement as they share the same main essential features \textcolor{black}{(see the review~\cite{Bush2021} for the comparison and relative merits of all these models). Here, we aim to simulate a gas of walkers which have long-range interactions. As the connectivity matrix of particles is fully coupled, the computational cost becomes a critical aspect}. As a result, we implement the walking drops dynamics by considering a \textcolor{black}{numerically optimized~\cite{Hubert2019,Hubert2021}} version of a discrete time evolution~\cite{Fort2010}, which can be summarized as follows.\\
\\
First, we model the wave created by a single impact on an unbounded fluid interface by a Bessel function of order $0$, $J_{0}$, centered at the point of impact. The standing wave frequency $f_{F}$ is directly linked to the vibration frequency of the bath $f_{0}$ by the relation $f_{F} = f_0/2$~\cite{Faraday1831,Rayleigh1883,Benjamin1954,Miles_1990,Douady1990,Kumar1994,Kumar1996,Protiere2006,Eddi2011,Molacek2013a}  
corresponding to a quasi-monochromatic wave at a wavelength $\lambda_F=6.1$~mm. As the drops impact the bath at constant relative phase every period $T_F=1/f_F=40$ ms, the wave field is \textcolor{black}{updated periodically.} Thus, a single impact at the surface position, $\bm{r}_0$, gives rise to a surface standing wave field with a spatial structure proportional to $J_0\left( k_F\Vert {\bm r}- {\bm r}_0\Vert \right)$, where ${\bm r}$ is a two-dimensional vector, and $k_F=2\pi/\lambda_F$ \textcolor{black}{the Faraday wave vector}. Experimentally, taking into account the fluid bath viscosity is known to lead to a spatial decay more pronounced than a Bessel function~\cite{Milewski2015,turton2018review,tadrist2018faraday}. \textcolor{black}{However considering a simplified Bessel wave field is a widely used approximation which has provided most of the theoretical predictions in the field}. As a consequence, we keep this simplified Bessel approximation for the wave field expression ~\cite{Labousse_Thesis,Bush2021}.\\
\\
Secondly, the persistence time of these waves is defined by \textcolor{black}{the memory time, $\mathrm{Me}\times T_F$, with \Me~the memory parameter}. The \textcolor{black}{larger} the value, the longer the waves remain on the surface and influence the dynamics of the drops. Experimentally, the memory time is tuned by varying the acceleration amplitude of the bath relative to the critical Faraday acceleration threshold. In the numerical model, the memory is a scalar parameter \Me~which we can vary at will. \Me~$\sim 10$ is considered as a low memory regime, and \Me~$\sim 50$ as a moderately small memory regime~\cite{Hubert2021}. We only investigate \Me~values ranging from 10 to 50. In practice, this memory range corresponds to an acceleration amplitude of the fluid bath $\gamma_m/\gamma_F=(1-1/\mathrm{Me})\in [0.9, 0.98] $, with $\gamma_F$ the fluid-dependent Faraday acceleration threshold. \\
\\
\begin{figure}
\centering
\includegraphics[width =\columnwidth]{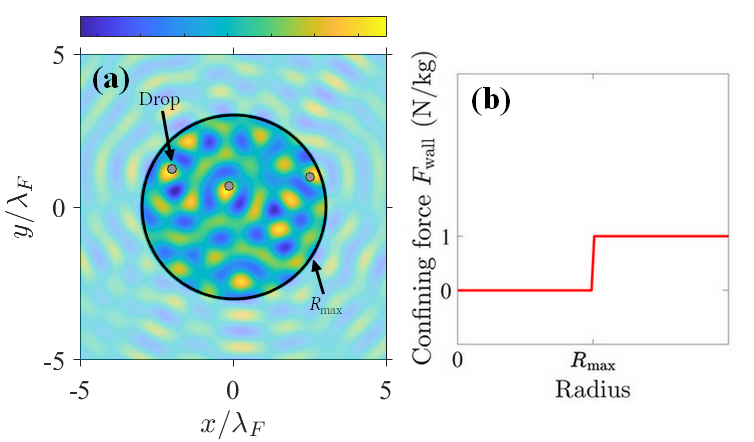}
\caption{Schematics of the system. (a) Example of the system with $N$ = 3 drops, \Me~= 30, $R_{\max} = 3\lambda_{F}$, seen from a top view. The coloured surface illustrates the wave field (in arbitrary units) created by the bouncing drops dotted in grey. The waves are not restrained by the outer wall, only the drops are. (b) A sketch of the confining force \textcolor{black}{per unit mass}: it is equivalent to a step of force of $1$~N\textcolor{black}{/kg} acting beyond a distance to the center larger than $R_{\max}$.}
\label{Fig:fig1}
\end{figure}
Thirdly, we consider the system as sketched in Fig.~\ref{Fig:fig1}. $N$ walking drops are confined by a very steep external potential acting \textcolor{black}{only} beyond a radius $R_{\max}$. The wave field is considered as not influenced by the presence of \textcolor{black}{this} confining potential and as a consequence, the waves are generated as if they were feeling an unbounded fluid interface. We \textcolor{black}{determine the evolution of} the position of the $i$-th drop ($i\in \left\{ 1,\ldots, N\right\}$) \textcolor{black}{from} time $t_p$ \textcolor{black}{and} at position $\bm{r}_i(t_p)$ to time $t_{p+1}=t_p+T_F$, i.e. one bounce further, \textcolor{black}{from the relation} $\bm{r}_i(t_{p+1})=\bm{r}_i(t_p)+\bm{v}(t_p)T_F$ and by computing \textcolor{black}{in parallel} the evolution of the velocity of the $i$-th drop, ${\bm v}_i$, as
\begin{equation}
    \begin{split}
    \bm{v}_i(t_{p+1}) = \bm{v}_i(t_p) + T_F\Bigl( -\eta \bm{v}_i(t_p) \\
   +~\bm{F}_{\mathrm{wave},i}(t_p) +\bm{F}_{\mathrm{wall},i}+\bm{F}_{\mathrm{drops}  \rightarrow \mathrm{drop} \;i}\Bigr).
    \label{Eq:eq1}
     \end{split}
\end{equation}
wherein $\eta$ is the dissipation coefficient is set to 4.72 s$
^{-1}$~\cite{Bacot2019,Labousse2016a}, which corresponds to \textcolor{black}{the situation of} a silicone oil with a kinematic viscosity of 20 cSt. It takes into account both the friction due to the lubrication layer between the drop and the bath, and the transfer of momentum to the fluid bath. 
$\bm{F}_{\mathrm{wave},i}$ is the force \textcolor{black}{per unit mass} propelling the $i$-th drop and is proportional to the local value of the two-dimensional gradient of the wave field. Specifically, it writes
\begin{equation}
\begin{split}
    \bm{F}_{\mathrm{wave},i}&(t_p) = -\frac{C T_F}{k_F} \bm{\nabla}_{\bm{r}_i(t_p)}\sum_{\substack{\mathrm{drop} \\  j=1 }}^N\biggl( \\
   & \sum_{k=p-3\mathrm{Me}}^{p}  J_{0}\left(k_{F}\Vert\bm{r}_i(t_p)-\bm{r}_{j}(t_k)\Vert\right)e^{\frac{-(p - k)}{\mathrm{Me}}}\biggr) 
\end{split}
\label{eqwave}
\end{equation}
The first sum accounts for the linear superposition of the wave generated by the drops $j$ ($j\in \left\{ 1\ldots N\right\}$). The second sum \textcolor{black}{originates from the memory kernel} which indicates that the amplitude of the waves generated by the $j$-th drop at time $t_k<t_p$ is exponentially damped. $C=1.1$ m.s$^{-3}$ is a coupling coefficient~\cite{Bacot2019,Labousse2016a} expected from a drop of diameter $0.9$~mm. Rigorously, the sum over $k$ should start from $k=0$ but thanks to the exponential term, we start the summation from $p-3$\Me, neglecting the other residual terms \textcolor{black}{(5 \%)}. \textcolor{black}{This truncation enables to reduce significantly the computational costs, without observable difference in the statistical convergence of the pair correlation function introduced further}.  $\bm{F}_{\mathrm{wall}}$ is the force \textcolor{black}{per unit mass} accounting for the presence of the outer wall. It is phenomenologically modelled by a step of force at \textcolor{black}{a radius} $R_{\max}$. It is only acting for a radius larger than $R_{\max}$ as sketched in Fig.~\ref{Fig:fig1}b. It writes
\begin{equation}
    \bm{F}_{\mathrm{wall}}=-F_0 \bm{e}_r H\left(r_i(t_p)-R_{\max}\right)
\end{equation}
with $\bm{e}_r$ the radial unit vector pointing outward the center, $H$ the Heaviside step function, $r_i=\Vert \bm{r}_i\Vert$ the distance of the $i$-th drop to the center and $F_0=1$~N/kg the wall force magnitude \textcolor{black}{per unit mass}. In practice, $\bm{F}_{\mathrm{wall}}$ is a step of force \textcolor{black}{whose magnitude in the external region} is at least one order of magnitude larger than all the other forces. Finally, $\bm{F}_{\mathrm{drops}  \rightarrow \mathrm{drop}\;i}$ accounts for the \textcolor{black}{short-range} drop-drop repulsion which we model as an elastic repulsion, with a spring constant (per unit mass) $K$, provided the distance between a pair of drops is smaller than twice the drop radius and zero otherwise. In what follows, we consider, $K=0$, i.e. $\bm{F}_{\mathrm{drops}  \rightarrow \mathrm{drop}\;i}={\bf 0}$, unless explicitly stated otherwise. This sum of forces determines the evolution of velocity of a given drop (Eq.~\ref{eqwave}) hence its position, period by period.\\
\\
The initial position of \textcolor{black}{the} drops is \textcolor{black}{generated} randomly and \textcolor{black}{uniformly} on a circular \textcolor{black}{area} of radius equal to half the radius of the whole accessible domain. We initialize the memory kernel by  generating randomly and uniformly the $3\times$\Me~last past positions around each drop within a distance 0.01$\times\lambda_F$ from the given drop. The drops evolve \textcolor{black}{along the fluid interface} for at least $10^4$ Faraday periods, \textcolor{black}{which is more than two orders of magnitude larger than the time required for a drop to move across the whole domain}. \textcolor{black}{Excepted for the cases with less than five drops, this simulation time is found to be sufficient to obtained a converged pair-correlation function as presented in Sec.~\ref{Sec:III}}. For \textcolor{black}{numerical simulation} with $5$ drops, $10^5$ Faraday periods are considered to decrease the measurements uncertainties. We discard the first $10^3$ Faraday periods to avoid any transient effects.

\section{Results : dynamical self-organization\label{Sec:III}}
\begin{figure}
\centering
\includegraphics[width =0.4\textwidth]{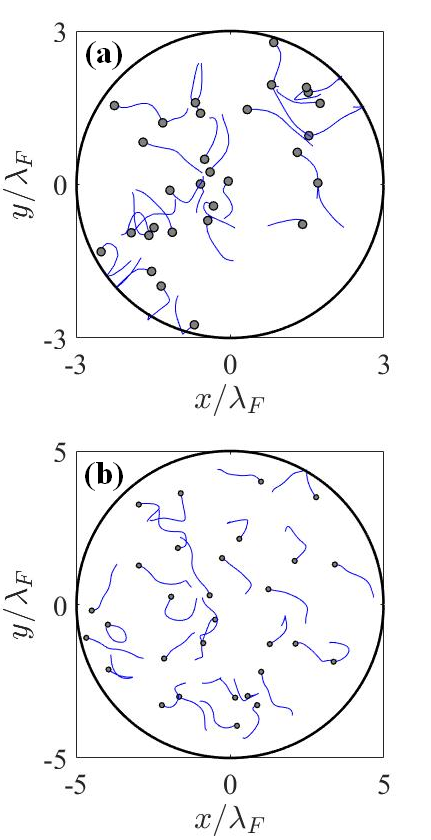}
\caption{Screenshots of $30$ walking drops evolving after $10^3$ time steps for a memory parameter \Me~= 30, with different bath sizes (a) $3\lambda_F$ and (b) $5\lambda_F$. The grey circles represent the drops with diameter $0.9$ mm. The blue lines link a drop to its \Me~past positions to illustrate the magnitude of the memory parameter. Two movies showing the drops moving during a few tens of periods are available in Supplemental Materials.}
\label{Fig:Fig2}
\end{figure}
We investigate the dynamics by changing the number of drops $N$, the domain radius $R_{\max}$, and the memory parameter \Me. Figures~\ref{Fig:Fig2}a and~\ref{Fig:Fig2}b illustrate two characteristic snapshots of the positions of 30 drops confined after $10^3$ time steps in a domain of radius $R_{\max}=3\lambda_F$ and $5\lambda_F$, respectively (see also Supplemental Movie 1 and 2). The magnitude of the memory parameter is indicated by a blue tail which visually links a drop to its \Me ~past positions. The mean density of drops, $\rho=N/(\pi R_{\max}^2)$ is low in both Fig.~\ref{Fig:Fig2}a and b, specifically $\rho_{\mathrm{drop}}\simeq 1.1~\lambda_F^{-2}$ and $0.4$~$\lambda_F^{-2}$, respectively, and we observe erratic trajectories, regardless of the bath size. We remark that sometimes, the drops walk together by forming small groups during a few tens of periods but we did not observe any reproducible patterns nor any obviously-identifiable long-standing stable emerging structure. Very small excursions outside $R_{\max}$ can be observed in Figures~\ref{Fig:Fig2} and are dictated by the precise details of the modelling of the wall repulsion. We also observe that the drops are overlaying and crossing each other sometimes. In this low drop density regime, we check that the dynamics is statistically not influenced by this effect \textcolor{black}{by comparing with the case when the force $\bm{F}_{\mathrm{drops} \rightarrow \mathrm{drop}\;i}$ is switched on}, an investigation which we shall discuss further. \\
\\
\begin{figure}
\includegraphics[width =0.38\textwidth]{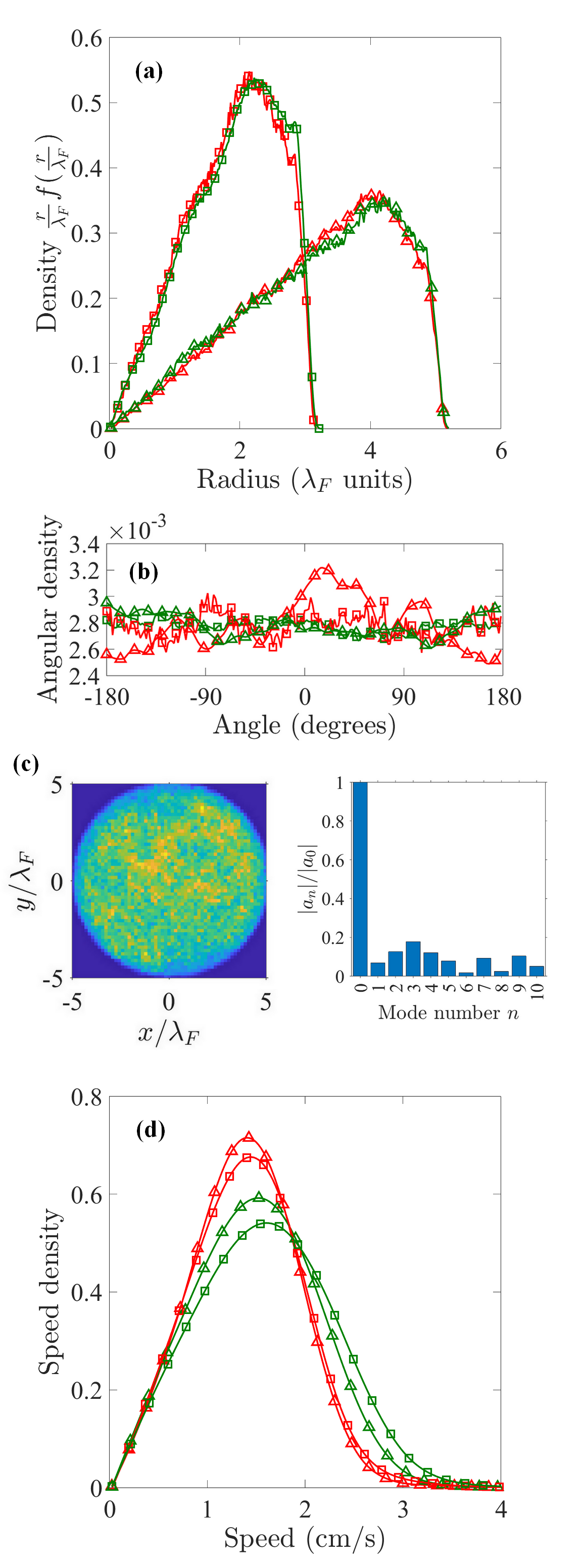}
\centering
\caption{Statistical properties of the dynamics (a) Radial, (b) angular probability density function (PDF). \textcolor{black}{(c) two-dimensional PDF for simulation time $10^4\; T_F$ and the relative absolute value of the weight modes $\vert a_n\vert /\vert a_0\vert$  of the mean wave field $\bar{h}$. (d)} speed PDF. Common color code: green $N$ = 30 and red $N$ = 5. The \textcolor{black}{squares} represent the situation $R_{\max}$ = 3$\lambda_{F}$ and the \textcolor{black}{triangles} $R_{\max}$ = 5$\lambda_{F}$. Me = 30 for all configurations.}
\label{Fig:Fig3}
\end{figure}
As we observe no apparent structure by looking \textcolor{black}{solely at the} dynamics, we proceed by investigating the statistical properties on the $N$ drops system \textcolor{black}{for a simulation time $10^4 \;T_F$}. We first compute the probability density function \textcolor{black}{(PDF)}, $\mathcal{F}(\bm{ r})$. We define the radial probability function, $f(r)$, by integrating $\mathcal{F}$ over all possible angular components, specifically, $f(r)=\int_{0}^{2\pi}\mathcal{F}(r,\theta)d\theta$. By construction, we get the normalization $\int f(r) rdr=\iint\mathcal{F}(r,\theta)rdrd\theta =1$. As a consequence, Figure~\ref{Fig:Fig3}a shows the evolution of $rf(r)$, with the length expressed in $\lambda_F$ units, for $N=5$ and $30$, a confining radius $R_{\max}=3\lambda_F$ and $5\lambda_F$, and for a common memory parameter \Me~= 30. For all  \textcolor{black}{parameter} configurations, we observe a linear dependency of $rf(r)$ with $r$ up to a radius $R^*\simeq R_{\max}-\lambda_F$, where $rf(r)$ is maximum. \textcolor{black}{For $r<R^*$, the slopes are independent of the drop density. Then, in this low density regime, we find that the slopes scale as $1/R_{\max}^2$. For distances from the center $r>R^*$, the drops feel the influence of the boundary and $rf(r)$ sharply decreases in the vicinity of $R_{\max}$. For radial distance to the center $r<R^*$, a linear behavior of $rf(r)$ means that $f$ is constant with $r$. So, in this region, the distribution of drops is radially homogenous. We note that this radial distribution is different from the case of a walker in a corral where both the walker motion and the wave modes are influenced by the boundary of the domain~\cite{Harris2013}. Additionally, we did not find any correlation between position and speed which is another difference with the corral system~\cite{Harris2013}.}\\
\\
Then, we analyse the angular structure of the system by computing the angular probability distribution $h(\theta)$, specifically $h(\theta)=\int\mathcal{F}(r,\theta)rdr$. Figure~\ref{Fig:Fig3}b shows that the drops angular distribution remains isotropic for all drop densities and all values of $R_{\max}$ investigated.\\
\\
\textcolor{black}{Figure~\ref{Fig:Fig3}c shows the 2D probability density function established for a simulation time $10^4 \;T_F$, 30 droplets and  \Me~$=30$. It does not have any particular structure and is homogenous up to some fluctuations which we assume to be due to the spatial coarse-graining. Then, we compute the convolution between the PDF and the Bessel wave kernel which gives the mean wave field, $\bar{h}$, according to the {\it mean wave field} theorem of Durey {et al.}~\cite{Durey2018}. We decompose $\bar{h}$ into a wave polar basis $(J_n(2\pi r/\lambda_F)e^{\mathrm{i}n\theta})_{n\in \mathbb{Z}}$ and calculate the weight modes $a_n$. The mean wave field is mainly dominated by the axisymmetric mode ($n=0$) which takes the spatial form of the Bessel function $J_0$. This is expected as the distribution of drops is found to be homogenous. So, because of this angular invariance, the only mode which contributes is the axisymmetric one. It is also a difference with~\cite{Hubert2021} where the mean wave field of a single particle, in asymptotically large memory limit, is not only a function of the density but requires the knowledge of higher-order temporal correlation functions. The comparison of the mean wave field with that obtained in~\cite{Tambasco2018} is interesting. In both cases, the spatial structure of the mean wave field is expected to be the same as the trajectories are statistically invariant by rotation. However, the prefactors must be very different as they encode the particular shape of the radial distributions which strongly differ in both cases.} \\
\\
Finally, we compute the speed probability distribution in Fig.~3d, from which we measure a mean speed $\bar{v}=1.40,~1.59,~1.38,~1.49$ cm/s, respectively, for $(N,R_{\max})= (5,3\lambda_F)$, $(30,3\lambda_F)$, $(5,5\lambda_F)$, $(30,5\lambda_F)$, respectively. The mean speeds are in the range expected experimentally. We observe that increasing the radius tends to \textcolor{black}{decrease the mean drops' speed}. Finally, we note that the systems with \textcolor{black}{larger} numbers of drops have more pronounced high-speed excursions. We investigate in more details this point in Figure 4 by plotting the speed density for \textcolor{black}{increasing number} of drops, up to 300.\\
\\
We rationalize the evolution of the speed \textcolor{black}{PDF} by taking inspirations from the active statistical theory developed in~\cite{Hubert2021}. In the large memory regime ($\mathrm{Me}\gg 10^2$) investigated by the authors, the motion of one single drop is equivalent to a self-propelled \textcolor{black}{particle coupled to a white noise thermal bath}. Following~\cite{Hubert2021}, we fit, in Fig.~\ref{Fig:Fig4}a, the numerical speed probability density functions with $h(v) =\alpha v\exp \left(-\beta ((v-v_0)^2) \right)$, wherein $v_0$ originates from a constrain on the speed arising from the self-propulsion of the drops and $\beta$ the effective temperature of the system. $\alpha$ is a normalisation factor such that $\int h(v) dv=1$. \textcolor{black}{Note that in the theoretical asymptotic limit $v_0 \rightarrow 0$,} $h$ identifies to a two-dimensional Maxwell-Boltzmann distribution as one would expect from a classical ideal gas in two dimensions. We observe some systematic quantitative differences between the numerical results and the fitting model especially with the tail of the distribution, suggesting strong remaining correlations. \\
\\
To reveal these correlations, we measure in Fig.~\ref{Fig:Fig4}b and ~\ref{Fig:Fig4}c, the evolution of the mean speed and of its standard deviation as a function of the number of drops along with their fits with 95$\%$ confidence intervals.  We observe a small and steady increase of the mean speed with the number of drops. It is a difference with the case of the promenade mode in which two correlated drops move slower than the case of one single drop. We conclude that we do not reach a statistical limit by increasing the number of drops, an important difference with an ideal gas and with the behavior of one single drop in the large memory regime~\cite{Hubert2021}. We note that the increase of the mean speed should obviously reach an upper limit in experiments as nonlinear wave effects are expected to start playing a saturation role. \\
\\
The speed fluctuations also increase with the number of drops (see Fig.~~\ref{Fig:Fig4}c) and we rationalize its evolution as follows. We expect that the squared speed fluctuation of a particle $i$, $\sigma_i^2$, should be the sum of two contributions: one intrinsic, denoted $\sigma_{i,0}^2$, and one due to the presence of all the other drops $j\neq i$, denoted $\Sigma_i^2$:
\begin{equation}
    \sigma_i^2=\sigma_{i,0}^2+\Sigma_i^2.
    \label{speed_fluci}
\end{equation}
In principle, $\Sigma_i^2$ is a function of ({\it i}) the number of drops, $N$, and ({\it ii}) all possible degrees of freedom of all the other drops $j\neq i$. All drops being indistinguishable, we assume that $\Sigma_i^2$ can be factorized as $\Sigma_i^2=\Sigma_{m,i}^2 F(N) $ where $F$ is a function independent on $i$ and only dependent on the number of drops, while $\Sigma_{m,i}^2$ is a function which depends, \textcolor{black}{in all generality}, on all possible degrees of freedom of all the other drops $j\neq i$. This factorization ansatz is equivalent to a meanfield approach, which we expect to hold for large $N$. \textcolor{black}{Additionally}, the number of drops is a scale-free parameter, so $F(N)$ is expected to be an algebraic function of $N$, specifically $F(N)=N^{\nu}$. $\nu$ is an exponent to be determined from the fitting onto the numerical results. By combining Eq.~\ref{speed_fluci} with the factorization ansatz, and then averaging on both time and on all drops $i=1 \ldots N$, we infer that the speed standard deviation $\delta v$ takes the form
\begin{equation}
    \delta v^2 = \delta v_0^2+ \delta v_m^2 N^{\nu}
    \label{Fluct_prediction}
\end{equation}
\textcolor{black}{with $\delta v_0$ and $\delta v_m$ two constants to be determined from the numerical results.
Fig.~4c indicates a good agreement with the proposed fit (Eq.~\ref{Fluct_prediction}).}\\
\\
\begin{figure}
\includegraphics[width =0.4\textwidth]{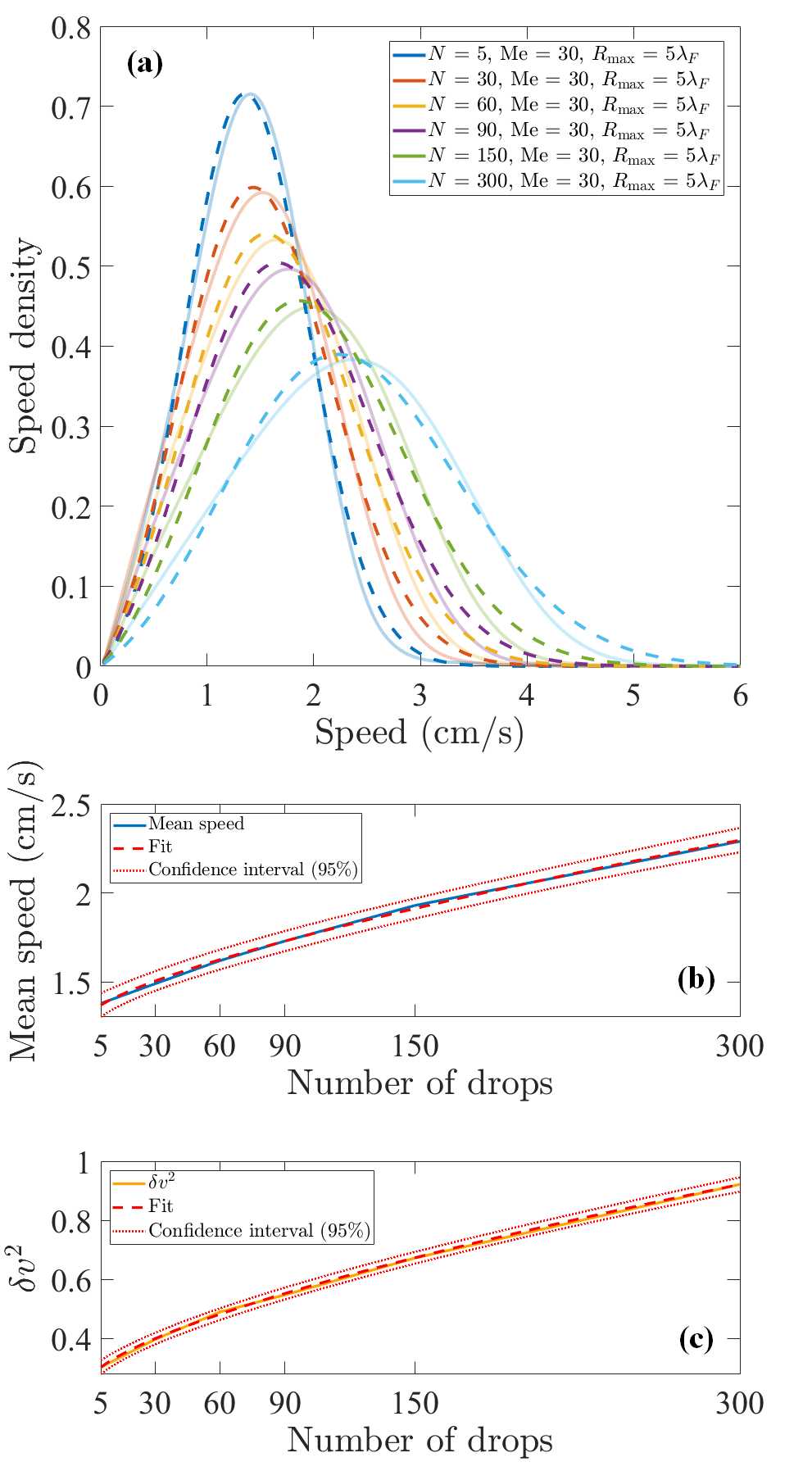}
\centering
\caption{Analysis of speed probability density functions. (a) Speed PDF for different numbers of drops, $\mathrm{Me}=30$ and $R_{\max}=5\lambda_F$. The dashed lines represent the fits by $h(v)=\alpha v\exp({-\beta (v-v_0)^2})$ of the numerical speed probability density functions (plain transparent lines). (b) Evolution of the mean speed as a function of the number of drops $N$. The fitting functions used is of the form $aN^b + c$ with $a = 0.0161$, $b = 0.7201$, $c = 0.3190$ and a determination coefficient $R^2= 0.9987$. (c) Evolution of the speed standard deviation as a function of the number of drops $N$. The fitting functions used is of the form of Eq. (5) with $\nu=0.681 $, $\delta v_0^2=1.36\; 10^{-2}$ cm${}^2$/s${}^2$, $\delta v_m^2=0.262$ cm${}^2$/s${}^2$ with a determination coefficient, $R^2=0.9996$.}
\label{Fig:Fig4}
\end{figure}

\begin{figure}
\includegraphics[width=0.4\textwidth]{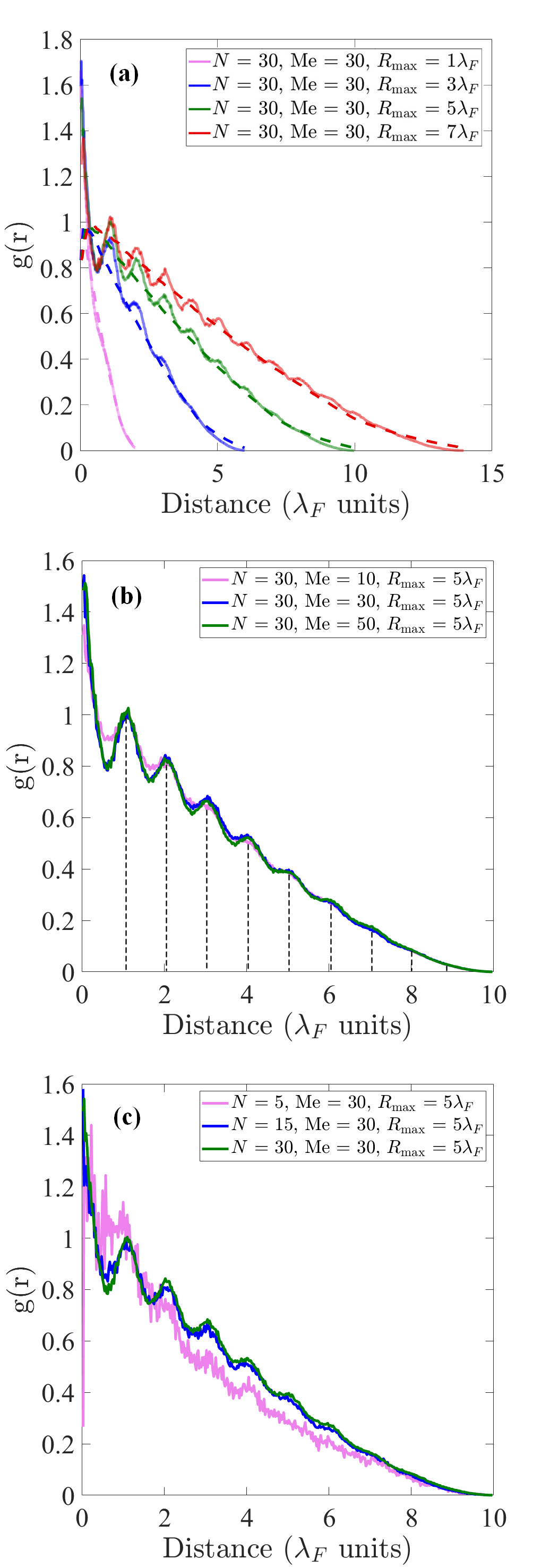}
\centering
\caption{Evolution of the pair correlation function $g$ for various (a) confining potential radii $R_{\max}$, (b) memory parameters Me, and (c) numbers of drops $N$. The dashed lines in (a) represent the theoretical situation for which all the drops where homogeneously distributed in a circular domain of radius $R_{\max}$. Vertical black lines in (b) represent the theoretical predictions for the position of the stable maxima (even solutions in Eq.~\ref{Eq:zeroJ1})}
\label{Fig:Fig5}
\end{figure}

\begin{figure}
\includegraphics[width=0.4\textwidth]{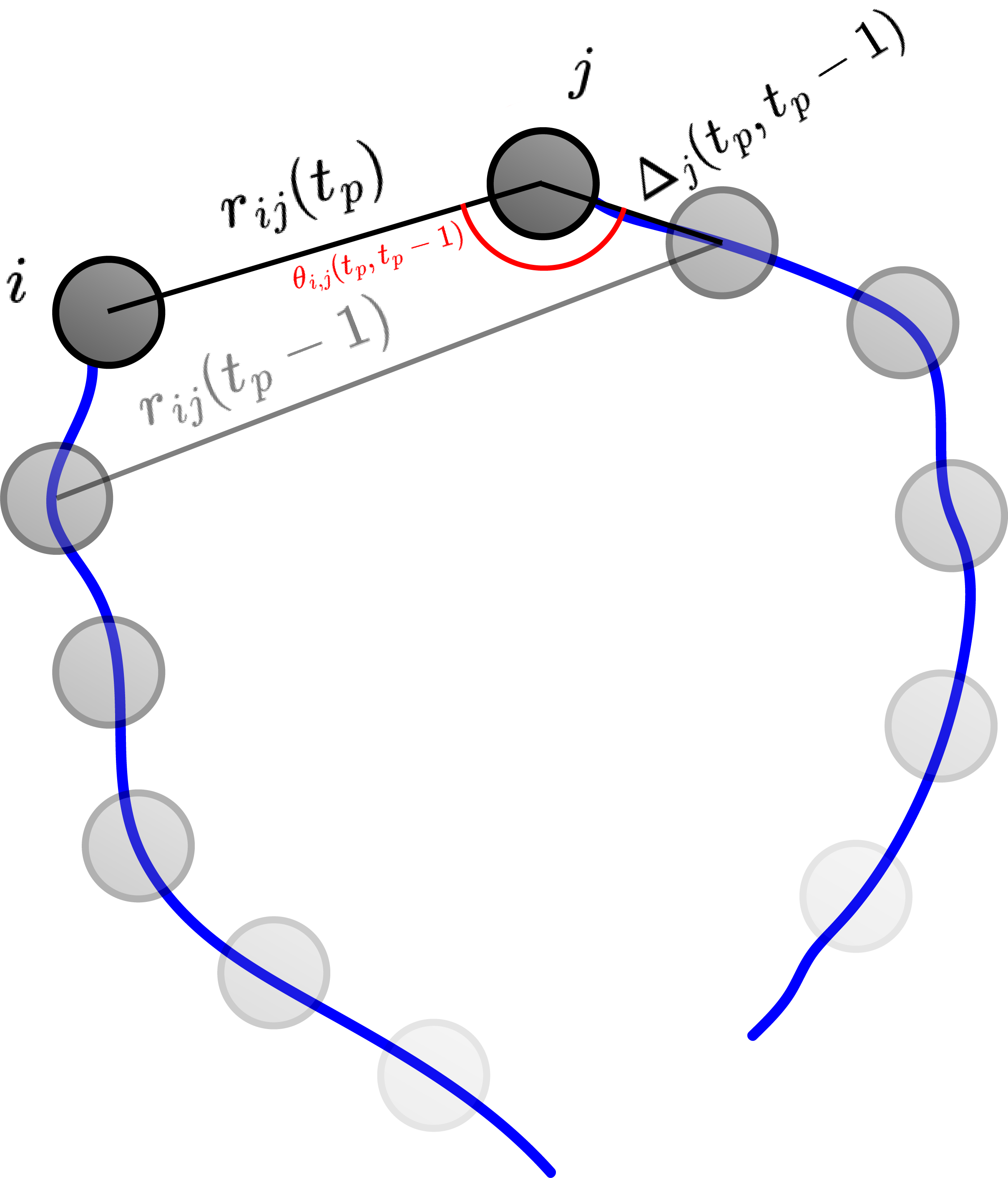}
\centering
\caption{Schematics to illustrate the notations used in Eq.~\ref{Graf}. The drops (dark grey circles) $i$ or $j$ ($i,j=1,\dots N$) follow their path marked in blue. Shaded grey circles indicate past positions.}
\label{Fig:Fig6}
\end{figure}

To investigate \textcolor{black}{these internal} correlations in more details, we aim at spotting the signature of any internal structure. To do so, we compute the drops pair correlation function
\begin{equation}
 g({\bm{r}})=\left\langle \sum_{i,j, i\neq j} \delta \left( {\bm r}-\left( {\bm r}_i-{\bm r}_j\right) \right)\right\rangle   
\end{equation}
where $\left\langle \ldots \right\rangle$ denotes an average over time. As there is no preferred \textcolor{black}{orientation}, we assume  $g({\bm{r}})= g({r}).$ The corresponding results are presented sequentially in Fig.~5, each panel corresponding to the variations of one single parameter. Fig.~\ref{Fig:Fig5}a indicates that \textcolor{black}{the pair correlation function decreases with the radial distance, up to twice $R_{\max}$, with the presence of well-pronounced equally-spaced peaks associated to preferred inter-drops distances. The number of observable peaks increases with the radius of the domain}. Peaks apart, the global decreasing trend is expected as a situation, for which the drops would be homogeneously distributed in a circular domain of radius $R_{\max}$, exhibit the same decreasing behavior. Fig.~\ref{Fig:Fig5}b indicates that varying memory in the range \Me$\;\in [10,50]$ does not change the pair correlation function profile, and moderately alter the magnitude of the peak-to-peak amplitude. Finally, Fig.~\ref{Fig:Fig5}c shows that varying the number of drops only affects the signal-to-noise ratio, which is very low for 5 drops. We observe on Figure~\ref{Fig:Fig5} local maxima on the curves. It means that drops tend to locate statistically from one to each other at a preferred set of quantized distances. Some of them are well pronounced with a well-measurable position and some of them are emerging which make their position difficult to estimate accurately. Among the peaks \textcolor{black}{which are} clearly formed, we measure, for example at $R_{\max} = 7\lambda_{F}$, the position of maxima $r^*/\lambda_F$= 1.09, 2.07, 3.08, 4.06, 5.04, 6.06, 6.97, 7.91 ($\pm$ 0.04). \\
\\
This quantized distance originates from the force due to the wave field (Eq.~\ref{eqwave}), which we rationalize now by considering the dominant terms in Eq.~\ref{eqwave}. We introduce the notations (\textcolor{black}{see also Fig.~\ref{Fig:Fig6}}): $\bm{r}_{ij}(t_p)=\bm{r}_i(t_p)-\bm{r}_{j}(t_p)$ \textcolor{black}{is a vector pointing from the drop $j$ to the drop $i$}, and $\bm{\Delta}_{j}(t_p,t_k)=\bm{r}_{j}(t_k)-\bm{r}_{j}(t_p)$ \textcolor{black}{linking the current position of the drop $j$ to one of its position in the past}. Additionally, we denote \textcolor{black}{the associated distances} $r_{i,j}(t_p)=\Vert \bm{r}_{ij}(t_p)\Vert$, $\Delta_{j}(t_p,t_k)=\Vert \bm{\Delta}_{j}(t_p,t_k)\Vert$, $J_n$ the Bessel function of first kind of order $n$ and $\theta_{i,j}(t_p,t_k)$ the relative angle between $\bm{r}_{ij}(t_p)$ and $\bm{\Delta}_{j}(t_p,t_k)$. Note that the symbol $\mathrm{i}$ in the complex exponential in Eq.~\ref{Graf} denotes the imaginary number and differs from the italicized subscript $i$. We identify the dominant terms in Eq.~\ref{eqwave} by proceeding as follows. As \textcolor{black}{varying} the memory \textcolor{black}{leaves} the position of maxima unchanged, we consider the regime of large \Me~which means that we neglect the exponential decay hereafter in Eq.~\ref{eqwave}. Then, we note that the wave term '$J_0$' in Eq.~\ref{eqwave} can be written
{\small
\begin{equation}
\begin{split}
  J_{0}(k_{F}&\Vert\bm{r}_i(t_p)-\bm{r}_{j}(t_k)\Vert  )\\
  & =J_{0}\left(k_{F}\Vert (\bm{r}_i(t_p)-\bm{r}_{j}(t_p))-(\bm{r}_{j}(t_k)-\bm{r}_{j}(t_p))\Vert\right)\\
  & =J_{0}\left(k_{F}\Vert\bm{r}_{ij}(t_p)-\bm{\Delta}_{j}(t_p,t_k)\Vert\right)\\
  & =\sum_{n=-\infty}^{+\infty}J_n(k_{F}r_{ij}(t_p))J_n(k_{F}\Delta_j(t_p,t_k))e^{\mathrm{i}n\theta_{i,j}(t_p,t_k)}
\end{split}
\label{Graf}
\end{equation}
}
In the first equality, we only added and subtracted $\bm{r}_{j}(t_p)$. To get the second equality, we use the notation defined above. Then, we use the Graf's addition theorem~\cite{NIST} to obtain the last equality. Then, to get the total wave field as in Eq.~\ref{eqwave}, we still have to perform a double summation over both the past positions \textcolor{black}{(indexed by $k$)} and drops pairs (\textcolor{black}{(indexed by $i$ and $j$)}). \textcolor{black}{Because of both this double summation and the erraticity of the drops' individual dynamics, the terms in Eq.~\ref{Graf} containing a complex phase rapidly self-average and give rise to small contributions to the wave field}. As a consequence, we expect the term $n=0$ in Eq.~\ref{Graf}, namely $U_{0,j}(r_{ij})=J_0(k_{F}r_{ij}(t_p))J_0(k_{F}\Delta_j(t_p,t_k))$, to be the leading term. As the wave-induced force, $\bm{F}_{\mathrm{wave},i}$ in Eq.~\ref{eqwave} derives from the gradient \textcolor{black}{of the wave field, we} expect the equilibrium solution to be given by the extrema of $U_{0,j}(r_{ij})$, apart from stability aspects. This can be found by looking for the zeros of $r \mapsto J_1(2\pi r/\lambda_F)$ which are 
\begin{equation}
    r^*/\lambda_F \approx 0,\; 0.63,\;  1.13,\; 1.63,\; 2.13,\; 2.63,\; 3.13,\; \ldots.
    \label{Eq:zeroJ1}
\end{equation} The even zeros correspond to those found numerically up to the uncertainty measurement (see vertical dashed lines in Fig.~5b). We recall that the situation is unstable and does not correspond to a crystalline structure. We conjecture that the odd zeros correspond to pair modes that are linearly unstable similarly to that observed in ~\cite{Couchman2019}. The peaks in the pair correlation function are the signature of a dynamically evolving internal structure and we have established its statistical wave origin.\\ 
\\

\begin{figure}
\includegraphics[width=0.4\textwidth]{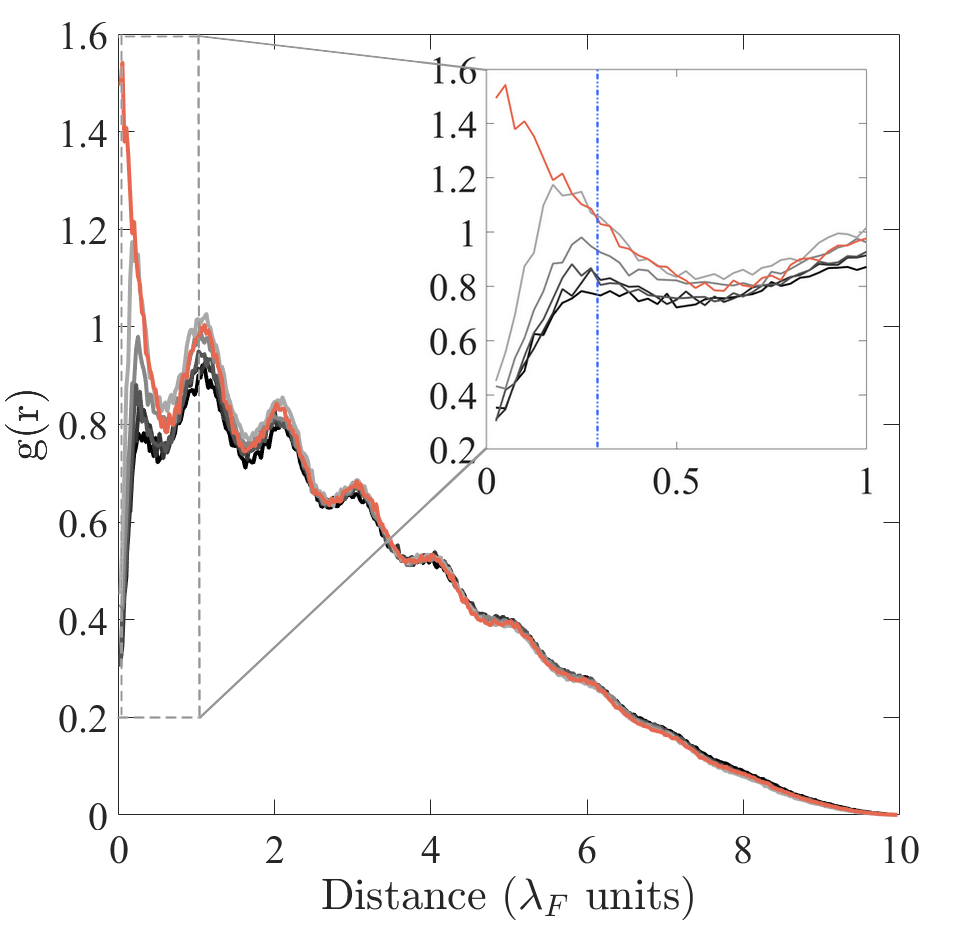}
\centering
\caption{Evolution of pairwise density function $g$ for the system $N$ = 30, Me = 30, $R_{\mathrm{max}}$ = 5$\lambda_F$ for various repulsion strengths. Color code:  from light grey to black, the spring constant per unit mass is $K=500,\; 1000,\; 1500,\; 2000\;, 2500$ s${}^{-1}$.kg${}^{-1}$. Darker curves correspond to higher repulsions. The salmon curve is the case in the absence of repulsion. Inset: zoom at low \textcolor{black}{inter-drops} distance. The blue vertical dashed line corresponds to the drop size in $\lambda_F$ units.}
\label{Fig:Fig7}
\end{figure}

Finally, we numerically predict that the pair correlation function exhibits a maximum at the origin. Experimentally, it is obviously impossible as short distance interactions must be taken into account. We present briefly in this last paragraph the influence of a short-distance elastic repulsion interaction. In Fig.~\ref{Fig:Fig7}, we observe that the systems with repulsion have pair correlation function $g(r)$ vanishing when $r$ tends to $0$, contrary to the system in the absence of repulsion. The stronger the repulsion, the more efficient the convergence towards $0$ at $r=0$, is. We also observe that including short distance interactions let the positions of the peaks, outside the origin, unchanged while their aspects are mainly unaltered.

\section{Conclusion and discussion\label{Sec:IV}}
We have numerically investigated the dynamics of a large number of walking drops that are confined by an external potential well while the surface Faraday waves are created as if they were in an unbounded fluid domain. The drops interact between each other by \textcolor{black}{the wave they collectively generate} and evolve in a complex chaotic dynamics which explores the whole accessible domain. A statistical analysis of the dynamics \textcolor{black}{shows that for a simulation time smaller than $10^4 \; T_F$ the drops probability density distribution is homogeneous up to a radial distance} approximately located at the edge $R_{\max}-\lambda_F$. A fine analysis of the pairwise \textcolor{black}{correlation} functions shows that while being dynamic, time-evolving and presenting many indications of a good mixing in the phase space, the system adopts, \textcolor{black}{on average,} preferred \textcolor{black}{inter-drops} distances whose origin has been rationalized by analysing the internal symmetry of the waves. Thus, this numerical investigation sheds light on a statistical many-body wave self-organization in an apparent erratic dynamics. \\ 
\\
\textcolor{black}{It is relatively surprising that the internal order evidenced in the pair correlation function do not lead to any identifiable signature in the position probability density function as it was observed in corral experiments~\cite{Harris2013}. However, we remain in a regime of parameters with a low density of drops, low memory while the simulation time, even if already reasonably long, is currently restricted by computational costs. Especially developing a more parallelizable numerical scheme, by taking inspiration from the field of molecular dynamics, would give access to both larger system sizes and longer simulation times for which the gas phases of walkers may not necessarily remain homogeneous. For example, in these regimes, can we observe an ordering of the gas phase or a condensate? The exploration of these still hypothetical  regimes is the natural next step.}
\section{Appendix}
The C++ numerical code to adapt to various platforms with a Readme file is available at the link repository https://mycore.core-cloud.net/index.php/s/qlFPCxcMHDEjQr6 .
\\
Supplementary Movie 1: movie associated to Fig. 2a.  Supplementary Movie 2: movie associated to Fig. 2b.


\bibliography{Bibliography_Helias_2022}

\begin{thebibliography}{72}%
\makeatletter
\providecommand \@ifxundefined [1]{%
 \@ifx{#1\undefined}
}%
\providecommand \@ifnum [1]{%
 \ifnum #1\expandafter \@firstoftwo
 \else \expandafter \@secondoftwo
 \fi
}%
\providecommand \@ifx [1]{%
 \ifx #1\expandafter \@firstoftwo
 \else \expandafter \@secondoftwo
 \fi
}%
\providecommand \natexlab [1]{#1}%
\providecommand \enquote  [1]{``#1''}%
\providecommand \bibnamefont  [1]{#1}%
\providecommand \bibfnamefont [1]{#1}%
\providecommand \citenamefont [1]{#1}%
\providecommand \href@noop [0]{\@secondoftwo}%
\providecommand \href [0]{\begingroup \@sanitize@url \@href}%
\providecommand \@href[1]{\@@startlink{#1}\@@href}%
\providecommand \@@href[1]{\endgroup#1\@@endlink}%
\providecommand \@sanitize@url [0]{\catcode `\\12\catcode `\$12\catcode
  `\&12\catcode `\#12\catcode `\^12\catcode `\_12\catcode `\%12\relax}%
\providecommand \@@startlink[1]{}%
\providecommand \@@endlink[0]{}%
\providecommand \url  [0]{\begingroup\@sanitize@url \@url }%
\providecommand \@url [1]{\endgroup\@href {#1}{\urlprefix }}%
\providecommand \urlprefix  [0]{URL }%
\providecommand \Eprint [0]{\href }%
\providecommand \doibase [0]{http://dx.doi.org/}%
\providecommand \selectlanguage [0]{\@gobble}%
\providecommand \bibinfo  [0]{\@secondoftwo}%
\providecommand \bibfield  [0]{\@secondoftwo}%
\providecommand \translation [1]{[#1]}%
\providecommand \BibitemOpen [0]{}%
\providecommand \bibitemStop [0]{}%
\providecommand \bibitemNoStop [0]{.\EOS\space}%
\providecommand \EOS [0]{\spacefactor3000\relax}%
\providecommand \BibitemShut  [1]{\csname bibitem#1\endcsname}%
\let\auto@bib@innerbib\@empty
\bibitem [{\citenamefont {Couder}\ \emph
  {et~al.}(2005{\natexlab{a}})\citenamefont {Couder}, \citenamefont
  {Proti{\`{e}}re}, \citenamefont {Fort},\ and\ \citenamefont
  {Boudaoud}}]{Couder2005}%
  \BibitemOpen
  \bibfield  {author} {\bibinfo {author} {\bibfnamefont {Y.}~\bibnamefont
  {Couder}}, \bibinfo {author} {\bibfnamefont {S.}~\bibnamefont
  {Proti{\`{e}}re}}, \bibinfo {author} {\bibfnamefont {E.}~\bibnamefont
  {Fort}}, \ and\ \bibinfo {author} {\bibfnamefont {A.}~\bibnamefont
  {Boudaoud}},\ }\href {\doibase 10.1038/437208a} {\bibfield  {journal}
  {\bibinfo  {journal} {Nature}\ }\textbf {\bibinfo {volume} {437}},\ \bibinfo
  {pages} {208} (\bibinfo {year} {2005}{\natexlab{a}})}\BibitemShut {NoStop}%
\bibitem [{\citenamefont {Bush}(2015)}]{Bush2015}%
  \BibitemOpen
  \bibfield  {author} {\bibinfo {author} {\bibfnamefont {J.~W.~M.}\
  \bibnamefont {Bush}},\ }\href {\doibase 10.1146/annurev-fluid-010814-014506}
  {\bibfield  {journal} {\bibinfo  {journal} {Annual Review of Fluid
  Mechanics}\ }\textbf {\bibinfo {volume} {47}},\ \bibinfo {pages} {269}
  (\bibinfo {year} {2015})}\BibitemShut {NoStop}%
\bibitem [{\citenamefont {Bush}\ and\ \citenamefont {Oza}(2021)}]{Bush2021}%
  \BibitemOpen
  \bibfield  {author} {\bibinfo {author} {\bibfnamefont {J.~W.~M.}\
  \bibnamefont {Bush}}\ and\ \bibinfo {author} {\bibfnamefont {A.~U.}\
  \bibnamefont {Oza}},\ }\href@noop {} {\bibfield  {journal} {\bibinfo
  {journal} {Report on Progress in Physics}\ }\textbf {\bibinfo {volume}
  {84}},\ \bibinfo {pages} {017001} (\bibinfo {year} {2021})}\BibitemShut
  {NoStop}%
\bibitem [{\citenamefont {Walker}(1978)}]{JWalker}%
  \BibitemOpen
  \bibfield  {author} {\bibinfo {author} {\bibfnamefont {J.}~\bibnamefont
  {Walker}},\ }\href@noop {} {\bibfield  {journal} {\bibinfo  {journal}
  {Scientific American}\ }\textbf {\bibinfo {volume} {238}},\ \bibinfo {pages}
  {151} (\bibinfo {year} {1978})}\BibitemShut {NoStop}%
\bibitem [{\citenamefont {Protiere}\ \emph {et~al.}(2006)\citenamefont
  {Protiere}, \citenamefont {Boudaoud},\ and\ \citenamefont
  {Couder}}]{JFM_Suzie}%
  \BibitemOpen
  \bibfield  {author} {\bibinfo {author} {\bibfnamefont {S.}~\bibnamefont
  {Protiere}}, \bibinfo {author} {\bibfnamefont {A.}~\bibnamefont {Boudaoud}},
  \ and\ \bibinfo {author} {\bibfnamefont {Y.}~\bibnamefont {Couder}},\
  }\href@noop {} {\bibfield  {journal} {\bibinfo  {journal} {Journal of Fluid
  Mechanics}\ }\textbf {\bibinfo {volume} {554}},\ \bibinfo {pages} {85}
  (\bibinfo {year} {2006})}\BibitemShut {NoStop}%
\bibitem [{\citenamefont {Couder}\ \emph
  {et~al.}(2005{\natexlab{b}})\citenamefont {Couder}, \citenamefont {Fort},
  \citenamefont {Gautier},\ and\ \citenamefont
  {Boudaoud}}]{from_bouncing_to_floating}%
  \BibitemOpen
  \bibfield  {author} {\bibinfo {author} {\bibfnamefont {Y.}~\bibnamefont
  {Couder}}, \bibinfo {author} {\bibfnamefont {E.}~\bibnamefont {Fort}},
  \bibinfo {author} {\bibfnamefont {C.}~\bibnamefont {Gautier}}, \ and\
  \bibinfo {author} {\bibfnamefont {A.}~\bibnamefont {Boudaoud}},\ }\href@noop
  {} {\bibfield  {journal} {\bibinfo  {journal} {Physical Review Letters}\
  }\textbf {\bibinfo {volume} {94}},\ \bibinfo {pages} {177801} (\bibinfo
  {year} {2005}{\natexlab{b}})}\BibitemShut {NoStop}%
\bibitem [{\citenamefont {Vandewalle}\ \emph {et~al.}(2008)\citenamefont
  {Vandewalle}, \citenamefont {Terwagne}, \citenamefont {Mulleners},
  \citenamefont {Gilet},\ and\ \citenamefont {Dorbolo}}]{Vandewalle_PRL}%
  \BibitemOpen
  \bibfield  {author} {\bibinfo {author} {\bibfnamefont {N.}~\bibnamefont
  {Vandewalle}}, \bibinfo {author} {\bibfnamefont {D.}~\bibnamefont
  {Terwagne}}, \bibinfo {author} {\bibfnamefont {K.}~\bibnamefont {Mulleners}},
  \bibinfo {author} {\bibfnamefont {T.}~\bibnamefont {Gilet}}, \ and\ \bibinfo
  {author} {\bibfnamefont {S.}~\bibnamefont {Dorbolo}},\ }\href@noop {}
  {\bibfield  {journal} {\bibinfo  {journal} {Physical Review Letters}\
  }\textbf {\bibinfo {volume} {100}} (\bibinfo {year} {2008})}\BibitemShut
  {NoStop}%
\bibitem [{\citenamefont {Eddi}\ \emph
  {et~al.}(2011{\natexlab{a}})\citenamefont {Eddi}, \citenamefont {Sultan},
  \citenamefont {Moukhtar}, \citenamefont {Fort}, \citenamefont {Rossi},\ and\
  \citenamefont {Couder}}]{Eddi2011}%
  \BibitemOpen
  \bibfield  {author} {\bibinfo {author} {\bibfnamefont {A.}~\bibnamefont
  {Eddi}}, \bibinfo {author} {\bibfnamefont {E.}~\bibnamefont {Sultan}},
  \bibinfo {author} {\bibfnamefont {J.}~\bibnamefont {Moukhtar}}, \bibinfo
  {author} {\bibfnamefont {E.}~\bibnamefont {Fort}}, \bibinfo {author}
  {\bibfnamefont {M.}~\bibnamefont {Rossi}}, \ and\ \bibinfo {author}
  {\bibfnamefont {Y.}~\bibnamefont {Couder}},\ }\href {\doibase
  10.1017/S0022112011000176} {\bibfield  {journal} {\bibinfo  {journal}
  {Journal of Fluid Mechanics}\ }\textbf {\bibinfo {volume} {674}},\ \bibinfo
  {pages} {433} (\bibinfo {year} {2011}{\natexlab{a}})}\BibitemShut {NoStop}%
\bibitem [{\citenamefont {Mol{\'{a}}{\v{c}}ek}\ and\ \citenamefont
  {Bush}(2013{\natexlab{a}})}]{Molacek2013a}%
  \BibitemOpen
  \bibfield  {author} {\bibinfo {author} {\bibfnamefont {J.}~\bibnamefont
  {Mol{\'{a}}{\v{c}}ek}}\ and\ \bibinfo {author} {\bibfnamefont {J.~W.~M.}\
  \bibnamefont {Bush}},\ }\href {\doibase 10.1017/jfm.2013.280} {\bibfield
  {journal} {\bibinfo  {journal} {Journal of Fluid Mechanics}\ }\textbf
  {\bibinfo {volume} {727}},\ \bibinfo {pages} {612} (\bibinfo {year}
  {2013}{\natexlab{a}})}\BibitemShut {NoStop}%
\bibitem [{\citenamefont {Mol{\'{a}}{\v{c}}ek}\ and\ \citenamefont
  {Bush}(2013{\natexlab{b}})}]{Molacek2013b}%
  \BibitemOpen
  \bibfield  {author} {\bibinfo {author} {\bibfnamefont {J.}~\bibnamefont
  {Mol{\'{a}}{\v{c}}ek}}\ and\ \bibinfo {author} {\bibfnamefont {J.~W.~M.}\
  \bibnamefont {Bush}},\ }\href {\doibase 10.1017/jfm.2013.279} {\bibfield
  {journal} {\bibinfo  {journal} {Journal of Fluid Mechanics}\ }\textbf
  {\bibinfo {volume} {727}},\ \bibinfo {pages} {582} (\bibinfo {year}
  {2013}{\natexlab{b}})}\BibitemShut {NoStop}%
\bibitem [{\citenamefont {Oza}\ \emph {et~al.}(2013)\citenamefont {Oza},
  \citenamefont {Rosales},\ and\ \citenamefont {Bush}}]{Oza_JFM_1_2013}%
  \BibitemOpen
  \bibfield  {author} {\bibinfo {author} {\bibfnamefont {A.~U.}\ \bibnamefont
  {Oza}}, \bibinfo {author} {\bibfnamefont {R.~R.}\ \bibnamefont {Rosales}}, \
  and\ \bibinfo {author} {\bibfnamefont {J.~W.~M.}\ \bibnamefont {Bush}},\
  }\href@noop {} {\bibfield  {journal} {\bibinfo  {journal} {Journal of Fluid
  Mechanics}\ }\textbf {\bibinfo {volume} {737}},\ \bibinfo {pages} {552}
  (\bibinfo {year} {2013})}\BibitemShut {NoStop}%
\bibitem [{\citenamefont {Milewski}\ \emph {et~al.}(2015)\citenamefont
  {Milewski}, \citenamefont {Galeano-Rios}, \citenamefont {Nachbin},\ and\
  \citenamefont {Bush}}]{Milewski2015}%
  \BibitemOpen
  \bibfield  {author} {\bibinfo {author} {\bibfnamefont {P.~A.}\ \bibnamefont
  {Milewski}}, \bibinfo {author} {\bibfnamefont {C.~A.}\ \bibnamefont
  {Galeano-Rios}}, \bibinfo {author} {\bibfnamefont {A.}~\bibnamefont
  {Nachbin}}, \ and\ \bibinfo {author} {\bibfnamefont {J.~W.~M.}\ \bibnamefont
  {Bush}},\ }\href {\doibase 10.1017/jfm.2015.386} {\bibfield  {journal}
  {\bibinfo  {journal} {Journal of Fluid Mechanics}\ }\textbf {\bibinfo
  {volume} {778}},\ \bibinfo {pages} {361} (\bibinfo {year}
  {2015})}\BibitemShut {NoStop}%
\bibitem [{\citenamefont {Durey}\ and\ \citenamefont
  {Milewski}(2017)}]{Durey2017}%
  \BibitemOpen
  \bibfield  {author} {\bibinfo {author} {\bibfnamefont {M.}~\bibnamefont
  {Durey}}\ and\ \bibinfo {author} {\bibfnamefont {P.~A.}\ \bibnamefont
  {Milewski}},\ }\href {\doibase 10.1017/jfm.2017.235} {\bibfield  {journal}
  {\bibinfo  {journal} {Journal of Fluid Mechanics}\ }\textbf {\bibinfo
  {volume} {821}},\ \bibinfo {pages} {296} (\bibinfo {year}
  {2017})}\BibitemShut {NoStop}%
\bibitem [{\citenamefont {Faraday}(1831)}]{Faraday1831}%
  \BibitemOpen
  \bibfield  {author} {\bibinfo {author} {\bibfnamefont {M.}~\bibnamefont
  {Faraday}},\ }\href {http://www.jstor.org/stable/107936} {\bibfield
  {journal} {\bibinfo  {journal} {Philosophical Transactions of the Royal
  Society of London}\ }\textbf {\bibinfo {volume} {121}},\ \bibinfo {pages}
  {299} (\bibinfo {year} {1831})}\BibitemShut {NoStop}%
\bibitem [{\citenamefont {Miles}\ and\ \citenamefont
  {Henderson}(1990)}]{Miles_1990}%
  \BibitemOpen
  \bibfield  {author} {\bibinfo {author} {\bibfnamefont {J.}~\bibnamefont
  {Miles}}\ and\ \bibinfo {author} {\bibfnamefont {D.}~\bibnamefont
  {Henderson}},\ }\href@noop {} {\bibfield  {journal} {\bibinfo  {journal}
  {Annual Review of Fluid Mechanics}\ }\textbf {\bibinfo {volume} {22}},\
  \bibinfo {pages} {143} (\bibinfo {year} {1990})}\BibitemShut {NoStop}%
\bibitem [{\citenamefont {Benjamin}\ and\ \citenamefont
  {Ursell}(1954)}]{Benjamin1954}%
  \BibitemOpen
  \bibfield  {author} {\bibinfo {author} {\bibfnamefont {T.~B.}\ \bibnamefont
  {Benjamin}}\ and\ \bibinfo {author} {\bibfnamefont {F.}~\bibnamefont
  {Ursell}},\ }\href@noop {} {\bibfield  {journal} {\bibinfo  {journal}
  {Proceedings of the Royal Society London A}\ }\textbf {\bibinfo {volume}
  {225}},\ \bibinfo {pages} {505} (\bibinfo {year} {1954})}\BibitemShut
  {NoStop}%
\bibitem [{\citenamefont {Couder}\ and\ \citenamefont
  {Fort}(2006)}]{Couder_Diffraction}%
  \BibitemOpen
  \bibfield  {author} {\bibinfo {author} {\bibfnamefont {Y.}~\bibnamefont
  {Couder}}\ and\ \bibinfo {author} {\bibfnamefont {E.}~\bibnamefont {Fort}},\
  }\href@noop {} {\bibfield  {journal} {\bibinfo  {journal} {Phys. Rev. Lett.}\
  }\textbf {\bibinfo {volume} {97}},\ \bibinfo {pages} {1} (\bibinfo {year}
  {2006})}\BibitemShut {NoStop}%
\bibitem [{\citenamefont {Pucci}\ \emph {et~al.}(2018)\citenamefont {Pucci},
  \citenamefont {Harris}, \citenamefont {Faria},\ and\ \citenamefont
  {Bush}}]{Pucci2018}%
  \BibitemOpen
  \bibfield  {author} {\bibinfo {author} {\bibfnamefont {G.}~\bibnamefont
  {Pucci}}, \bibinfo {author} {\bibfnamefont {D.~M.}\ \bibnamefont {Harris}},
  \bibinfo {author} {\bibfnamefont {L.~M.}\ \bibnamefont {Faria}}, \ and\
  \bibinfo {author} {\bibfnamefont {J.~W.~M.}\ \bibnamefont {Bush}},\ }\href
  {\doibase 10.1017/jfm.2017.790} {\bibfield  {journal} {\bibinfo  {journal}
  {Journal of Fluid Mechanics}\ }\textbf {\bibinfo {volume} {835}},\ \bibinfo
  {pages} {1136–1156} (\bibinfo {year} {2018})}\BibitemShut {NoStop}%
\bibitem [{\citenamefont {Ellegaard}\ and\ \citenamefont
  {Levinsen}(2020)}]{EllegaardLevinsen2020}%
  \BibitemOpen
  \bibfield  {author} {\bibinfo {author} {\bibfnamefont {C.}~\bibnamefont
  {Ellegaard}}\ and\ \bibinfo {author} {\bibfnamefont {M.~T.}\ \bibnamefont
  {Levinsen}},\ }\href {\doibase 10.1103/PhysRevE.102.023115} {\bibfield
  {journal} {\bibinfo  {journal} {Phys. Rev. E}\ }\textbf {\bibinfo {volume}
  {102}},\ \bibinfo {pages} {023115} (\bibinfo {year} {2020})}\BibitemShut
  {NoStop}%
\bibitem [{\citenamefont {Harris}\ \emph {et~al.}(2013)\citenamefont {Harris},
  \citenamefont {Moukhtar}, \citenamefont {Fort}, \citenamefont {Couder},\ and\
  \citenamefont {Bush}}]{Harris2013}%
  \BibitemOpen
  \bibfield  {author} {\bibinfo {author} {\bibfnamefont {D.~M.}\ \bibnamefont
  {Harris}}, \bibinfo {author} {\bibfnamefont {J.}~\bibnamefont {Moukhtar}},
  \bibinfo {author} {\bibfnamefont {E.}~\bibnamefont {Fort}}, \bibinfo {author}
  {\bibfnamefont {Y.}~\bibnamefont {Couder}}, \ and\ \bibinfo {author}
  {\bibfnamefont {J.~W.~M.}\ \bibnamefont {Bush}},\ }\href {\doibase
  10.1103/PhysRevE.88.011001} {\bibfield  {journal} {\bibinfo  {journal}
  {Physical Review E}\ }\textbf {\bibinfo {volume} {88}},\ \bibinfo {pages}
  {011001} (\bibinfo {year} {2013})}\BibitemShut {NoStop}%
\bibitem [{\citenamefont {Gilet}(2014)}]{Gilet2014}%
  \BibitemOpen
  \bibfield  {author} {\bibinfo {author} {\bibfnamefont {T.}~\bibnamefont
  {Gilet}},\ }\href {\doibase 10.1103/PhysRevE.90.052917} {\bibfield  {journal}
  {\bibinfo  {journal} {Physical Review E}\ }\textbf {\bibinfo {volume} {90}},\
  \bibinfo {pages} {052917} (\bibinfo {year} {2014})}\BibitemShut {NoStop}%
\bibitem [{\citenamefont {Gilet}(2016)}]{Gilet2016}%
  \BibitemOpen
  \bibfield  {author} {\bibinfo {author} {\bibfnamefont {T.}~\bibnamefont
  {Gilet}},\ }\href {\doibase 10.1103/PhysRevE.93.042202} {\bibfield  {journal}
  {\bibinfo  {journal} {Physical Review E}\ }\textbf {\bibinfo {volume} {93}}
  (\bibinfo {year} {2016}),\ 10.1103/PhysRevE.93.042202}\BibitemShut {NoStop}%
\bibitem [{\citenamefont {S{\'{a}}enz}\ \emph {et~al.}(2018)\citenamefont
  {S{\'{a}}enz}, \citenamefont {Cristea-Platon},\ and\ \citenamefont
  {Bush}}]{Saenz2018}%
  \BibitemOpen
  \bibfield  {author} {\bibinfo {author} {\bibfnamefont {P.~J.}\ \bibnamefont
  {S{\'{a}}enz}}, \bibinfo {author} {\bibfnamefont {T.}~\bibnamefont
  {Cristea-Platon}}, \ and\ \bibinfo {author} {\bibfnamefont {J.~W.~M.}\
  \bibnamefont {Bush}},\ }\href {\doibase 10.1038/s41567-017-0003-x} {\bibfield
   {journal} {\bibinfo  {journal} {Nature Physics}\ }\textbf {\bibinfo {volume}
  {14}},\ \bibinfo {pages} {315} (\bibinfo {year} {2018})}\BibitemShut
  {NoStop}%
\bibitem [{\citenamefont {Durey}\ \emph
  {et~al.}(2020{\natexlab{a}})\citenamefont {Durey}, \citenamefont {Milewski},\
  and\ \citenamefont {Wang}}]{durey_milewski_wang_2020}%
  \BibitemOpen
  \bibfield  {author} {\bibinfo {author} {\bibfnamefont {M.}~\bibnamefont
  {Durey}}, \bibinfo {author} {\bibfnamefont {P.~A.}\ \bibnamefont {Milewski}},
  \ and\ \bibinfo {author} {\bibfnamefont {Z.}~\bibnamefont {Wang}},\ }\href
  {\doibase 10.1017/jfm.2020.140} {\bibfield  {journal} {\bibinfo  {journal}
  {Journal of Fluid Mechanics}\ }\textbf {\bibinfo {volume} {891}},\ \bibinfo
  {pages} {A3} (\bibinfo {year} {2020}{\natexlab{a}})}\BibitemShut {NoStop}%
\bibitem [{\citenamefont {Fort}\ \emph {et~al.}(2010)\citenamefont {Fort},
  \citenamefont {Eddi}, \citenamefont {Boudaoud}, \citenamefont {Moukhtar},\
  and\ \citenamefont {Couder}}]{Fort2010}%
  \BibitemOpen
  \bibfield  {author} {\bibinfo {author} {\bibfnamefont {E.}~\bibnamefont
  {Fort}}, \bibinfo {author} {\bibfnamefont {A.}~\bibnamefont {Eddi}}, \bibinfo
  {author} {\bibfnamefont {A.}~\bibnamefont {Boudaoud}}, \bibinfo {author}
  {\bibfnamefont {J.}~\bibnamefont {Moukhtar}}, \ and\ \bibinfo {author}
  {\bibfnamefont {Y.}~\bibnamefont {Couder}},\ }\href {\doibase
  10.1073/pnas.1007386107} {\bibfield  {journal} {\bibinfo  {journal}
  {Proceedings of the National Academy of Sciences}\ }\textbf {\bibinfo
  {volume} {107}},\ \bibinfo {pages} {17515} (\bibinfo {year}
  {2010})}\BibitemShut {NoStop}%
\bibitem [{\citenamefont {Labousse}\ \emph {et~al.}(2016)\citenamefont
  {Labousse}, \citenamefont {Oza}, \citenamefont {Perrard},\ and\ \citenamefont
  {Bush}}]{Labousse2016a}%
  \BibitemOpen
  \bibfield  {author} {\bibinfo {author} {\bibfnamefont {M.}~\bibnamefont
  {Labousse}}, \bibinfo {author} {\bibfnamefont {A.~U.}\ \bibnamefont {Oza}},
  \bibinfo {author} {\bibfnamefont {S.}~\bibnamefont {Perrard}}, \ and\
  \bibinfo {author} {\bibfnamefont {J.~W.~M.}\ \bibnamefont {Bush}},\
  }\href@noop {} {\bibfield  {journal} {\bibinfo  {journal} {Physical Review
  E}\ }\textbf {\bibinfo {volume} {93}} (\bibinfo {year} {2016})}\BibitemShut
  {NoStop}%
\bibitem [{\citenamefont {Oza}\ \emph {et~al.}(2014)\citenamefont {Oza},
  \citenamefont {Harris}, \citenamefont {Rosales},\ and\ \citenamefont
  {Bush}}]{Oza2014}%
  \BibitemOpen
  \bibfield  {author} {\bibinfo {author} {\bibfnamefont {A.~U.}\ \bibnamefont
  {Oza}}, \bibinfo {author} {\bibfnamefont {D.~M.}\ \bibnamefont {Harris}},
  \bibinfo {author} {\bibfnamefont {R.~R.}\ \bibnamefont {Rosales}}, \ and\
  \bibinfo {author} {\bibfnamefont {J.~W.~M.}\ \bibnamefont {Bush}},\ }\href
  {\doibase 10.1017/jfm.2014.50} {\bibfield  {journal} {\bibinfo  {journal}
  {Journal of Fluid Mechanics}\ }\textbf {\bibinfo {volume} {744}},\ \bibinfo
  {pages} {404} (\bibinfo {year} {2014})}\BibitemShut {NoStop}%
\bibitem [{\citenamefont {Perrard}\ \emph
  {et~al.}(2014{\natexlab{a}})\citenamefont {Perrard}, \citenamefont
  {Labousse}, \citenamefont {Miskin}, \citenamefont {Fort},\ and\ \citenamefont
  {Couder}}]{Perrard2014a}%
  \BibitemOpen
  \bibfield  {author} {\bibinfo {author} {\bibfnamefont {S.}~\bibnamefont
  {Perrard}}, \bibinfo {author} {\bibfnamefont {M.}~\bibnamefont {Labousse}},
  \bibinfo {author} {\bibfnamefont {M.}~\bibnamefont {Miskin}}, \bibinfo
  {author} {\bibfnamefont {E.}~\bibnamefont {Fort}}, \ and\ \bibinfo {author}
  {\bibfnamefont {Y.}~\bibnamefont {Couder}},\ }\href {\doibase
  10.1038/ncomms4219} {\bibfield  {journal} {\bibinfo  {journal} {Nature
  Communications}\ }\textbf {\bibinfo {volume} {5}},\ \bibinfo {pages} {3219}
  (\bibinfo {year} {2014}{\natexlab{a}})}\BibitemShut {NoStop}%
\bibitem [{\citenamefont {Perrard}\ \emph
  {et~al.}(2014{\natexlab{b}})\citenamefont {Perrard}, \citenamefont
  {Labousse}, \citenamefont {Fort},\ and\ \citenamefont
  {Couder}}]{Perrard2014}%
  \BibitemOpen
  \bibfield  {author} {\bibinfo {author} {\bibfnamefont {S.}~\bibnamefont
  {Perrard}}, \bibinfo {author} {\bibfnamefont {M.}~\bibnamefont {Labousse}},
  \bibinfo {author} {\bibfnamefont {E.}~\bibnamefont {Fort}}, \ and\ \bibinfo
  {author} {\bibfnamefont {Y.}~\bibnamefont {Couder}},\ }\href {\doibase
  10.1103/PhysRevLett.113.104101} {\bibfield  {journal} {\bibinfo  {journal}
  {Physical Review Letters}\ }\textbf {\bibinfo {volume} {113}},\ \bibinfo
  {pages} {104101} (\bibinfo {year} {2014}{\natexlab{b}})}\BibitemShut
  {NoStop}%
\bibitem [{\citenamefont {Labousse}\ \emph {et~al.}(2014)\citenamefont
  {Labousse}, \citenamefont {Perrard}, \citenamefont {Couder},\ and\
  \citenamefont {Fort}}]{Labousse2014}%
  \BibitemOpen
  \bibfield  {author} {\bibinfo {author} {\bibfnamefont {M.}~\bibnamefont
  {Labousse}}, \bibinfo {author} {\bibfnamefont {S.}~\bibnamefont {Perrard}},
  \bibinfo {author} {\bibfnamefont {Y.}~\bibnamefont {Couder}}, \ and\ \bibinfo
  {author} {\bibfnamefont {E.}~\bibnamefont {Fort}},\ }\href {\doibase
  10.1088/1367-2630/16/11/113027} {\bibfield  {journal} {\bibinfo  {journal}
  {New Journal of Physics}\ }\textbf {\bibinfo {volume} {16}},\ \bibinfo
  {pages} {113027} (\bibinfo {year} {2014})}\BibitemShut {NoStop}%
\bibitem [{\citenamefont {S{\'a}enz}\ \emph {et~al.}(2020)\citenamefont
  {S{\'a}enz}, \citenamefont {Cristea-Platon},\ and\ \citenamefont
  {Bush}}]{Saenz2020_friedel}%
  \BibitemOpen
  \bibfield  {author} {\bibinfo {author} {\bibfnamefont {P.~J.}\ \bibnamefont
  {S{\'a}enz}}, \bibinfo {author} {\bibfnamefont {T.}~\bibnamefont
  {Cristea-Platon}}, \ and\ \bibinfo {author} {\bibfnamefont {J.~W.~M.}\
  \bibnamefont {Bush}},\ }\href {\doibase 10.1126/sciadv.aay9234} {\bibfield
  {journal} {\bibinfo  {journal} {Science Advances}\ }\textbf {\bibinfo
  {volume} {6}} (\bibinfo {year} {2020}),\ 10.1126/sciadv.aay9234}\BibitemShut
  {NoStop}%
\bibitem [{\citenamefont {Hubert}\ \emph {et~al.}(2019)\citenamefont {Hubert},
  \citenamefont {Perrard}, \citenamefont {Labousse}, \citenamefont
  {Vandewalle},\ and\ \citenamefont {Couder}}]{Hubert2019}%
  \BibitemOpen
  \bibfield  {author} {\bibinfo {author} {\bibfnamefont {M.}~\bibnamefont
  {Hubert}}, \bibinfo {author} {\bibfnamefont {S.}~\bibnamefont {Perrard}},
  \bibinfo {author} {\bibfnamefont {M.}~\bibnamefont {Labousse}}, \bibinfo
  {author} {\bibfnamefont {N.}~\bibnamefont {Vandewalle}}, \ and\ \bibinfo
  {author} {\bibfnamefont {Y.}~\bibnamefont {Couder}},\ }\href {\doibase
  10.1103/PhysRevE.100.032201} {\bibfield  {journal} {\bibinfo  {journal}
  {Physical Review E}\ }\textbf {\bibinfo {volume} {100}},\ \bibinfo {pages}
  {032201} (\bibinfo {year} {2019})}\BibitemShut {NoStop}%
\bibitem [{\citenamefont {Bacot}\ \emph {et~al.}(2019)\citenamefont {Bacot},
  \citenamefont {Perrard}, \citenamefont {Labousse}, \citenamefont {Couder},\
  and\ \citenamefont {Fort}}]{Bacot2019}%
  \BibitemOpen
  \bibfield  {author} {\bibinfo {author} {\bibfnamefont {V.}~\bibnamefont
  {Bacot}}, \bibinfo {author} {\bibfnamefont {S.}~\bibnamefont {Perrard}},
  \bibinfo {author} {\bibfnamefont {M.}~\bibnamefont {Labousse}}, \bibinfo
  {author} {\bibfnamefont {Y.}~\bibnamefont {Couder}}, \ and\ \bibinfo {author}
  {\bibfnamefont {E.}~\bibnamefont {Fort}},\ }\href {\doibase
  10.1103/PhysRevLett.122.104303} {\bibfield  {journal} {\bibinfo  {journal}
  {Physical Review Letters}\ }\textbf {\bibinfo {volume} {122}},\ \bibinfo
  {pages} {104303} (\bibinfo {year} {2019})}\BibitemShut {NoStop}%
\bibitem [{\citenamefont {Durey}\ \emph {et~al.}(2018)\citenamefont {Durey},
  \citenamefont {Milewski},\ and\ \citenamefont {Bush}}]{Durey2018}%
  \BibitemOpen
  \bibfield  {author} {\bibinfo {author} {\bibfnamefont {M.}~\bibnamefont
  {Durey}}, \bibinfo {author} {\bibfnamefont {P.~A.}\ \bibnamefont {Milewski}},
  \ and\ \bibinfo {author} {\bibfnamefont {J.~W.~M.}\ \bibnamefont {Bush}},\
  }\href {\doibase 10.1063/1.5030639} {\bibfield  {journal} {\bibinfo
  {journal} {Chaos: An Interdisciplinary Journal of Nonlinear Science}\
  }\textbf {\bibinfo {volume} {28}},\ \bibinfo {pages} {096108} (\bibinfo
  {year} {2018})}\BibitemShut {NoStop}%
\bibitem [{\citenamefont {Durey}(2020)}]{Durey2020}%
  \BibitemOpen
  \bibfield  {author} {\bibinfo {author} {\bibfnamefont {M.}~\bibnamefont
  {Durey}},\ }\href {\doibase 10.1063/5.0020775} {\bibfield  {journal}
  {\bibinfo  {journal} {Chaos: An Interdisciplinary Journal of Nonlinear
  Science}\ }\textbf {\bibinfo {volume} {30}},\ \bibinfo {pages} {103115}
  (\bibinfo {year} {2020})}\BibitemShut {NoStop}%
\bibitem [{\citenamefont {Durey}\ and\ \citenamefont
  {Bush}(2021)}]{Durey_chaos}%
  \BibitemOpen
  \bibfield  {author} {\bibinfo {author} {\bibfnamefont {M.}~\bibnamefont
  {Durey}}\ and\ \bibinfo {author} {\bibfnamefont {J.~W.~M.}\ \bibnamefont
  {Bush}},\ }\href@noop {} {\bibfield  {journal} {\bibinfo  {journal} {Chaos:
  An Interdisciplinary Journal of Nonlinear Science}\ }\textbf {\bibinfo
  {volume} {31}},\ \bibinfo {pages} {033136} (\bibinfo {year}
  {2021})}\BibitemShut {NoStop}%
\bibitem [{\citenamefont {Durey}\ \emph
  {et~al.}(2020{\natexlab{b}})\citenamefont {Durey}, \citenamefont {Turton},\
  and\ \citenamefont {Bush}}]{Durey_oscillation}%
  \BibitemOpen
  \bibfield  {author} {\bibinfo {author} {\bibfnamefont {M.}~\bibnamefont
  {Durey}}, \bibinfo {author} {\bibfnamefont {S.}~\bibnamefont {Turton}}, \
  and\ \bibinfo {author} {\bibfnamefont {J.~W.~M.}\ \bibnamefont {Bush}},\
  }\href@noop {} {\bibfield  {journal} {\bibinfo  {journal} {Proceedings of the
  Royal Society A}\ }\textbf {\bibinfo {volume} {476}},\ \bibinfo {pages}
  {2239} (\bibinfo {year} {2020}{\natexlab{b}})}\BibitemShut {NoStop}%
\bibitem [{\citenamefont {Devauchelle}\ \emph {et~al.}(2020)\citenamefont
  {Devauchelle}, \citenamefont {Lajeunesse}, \citenamefont {James},
  \citenamefont {Josserand},\ and\ \citenamefont {Lagr\'{e}e}}]{Devauchelle}%
  \BibitemOpen
  \bibfield  {author} {\bibinfo {author} {\bibfnamefont {O.}~\bibnamefont
  {Devauchelle}}, \bibinfo {author} {\bibfnamefont {E.}~\bibnamefont
  {Lajeunesse}}, \bibinfo {author} {\bibfnamefont {F.}~\bibnamefont {James}},
  \bibinfo {author} {\bibfnamefont {C.}~\bibnamefont {Josserand}}, \ and\
  \bibinfo {author} {\bibfnamefont {P.}~\bibnamefont {Lagr\'{e}e}},\ }\href
  {\doibase 10.5802/crmeca.25} {\bibfield  {journal} {\bibinfo  {journal}
  {Comptes Rendus. M\'ecanique}\ }\textbf {\bibinfo {volume} {438}},\ \bibinfo
  {pages} {591} (\bibinfo {year} {2020})}\BibitemShut {NoStop}%
\bibitem [{\citenamefont {Valani}\ \emph {et~al.}(2021)\citenamefont {Valani},
  \citenamefont {Slim}, \citenamefont {Paganin}, \citenamefont {Simula},\ and\
  \citenamefont {Vo}}]{Rahil}%
  \BibitemOpen
  \bibfield  {author} {\bibinfo {author} {\bibfnamefont {R.~N.}\ \bibnamefont
  {Valani}}, \bibinfo {author} {\bibfnamefont {A.~C.}\ \bibnamefont {Slim}},
  \bibinfo {author} {\bibfnamefont {D.~M.}\ \bibnamefont {Paganin}}, \bibinfo
  {author} {\bibfnamefont {T.~P.}\ \bibnamefont {Simula}}, \ and\ \bibinfo
  {author} {\bibfnamefont {T.}~\bibnamefont {Vo}},\ }\href {\doibase
  10.1103/PhysRevE.104.015106} {\bibfield  {journal} {\bibinfo  {journal}
  {Physical Review E}\ }\textbf {\bibinfo {volume} {104}},\ \bibinfo {pages}
  {015106} (\bibinfo {year} {2021})}\BibitemShut {NoStop}%
\bibitem [{\citenamefont {Hubert}\ \emph {et~al.}(2022)\citenamefont {Hubert},
  \citenamefont {Perrard}, \citenamefont {Vandewalle},\ and\ \citenamefont
  {Labousse}}]{Hubert2021}%
  \BibitemOpen
  \bibfield  {author} {\bibinfo {author} {\bibfnamefont {M.}~\bibnamefont
  {Hubert}}, \bibinfo {author} {\bibfnamefont {S.}~\bibnamefont {Perrard}},
  \bibinfo {author} {\bibfnamefont {N.}~\bibnamefont {Vandewalle}}, \ and\
  \bibinfo {author} {\bibfnamefont {M.}~\bibnamefont {Labousse}},\ }\href@noop
  {} {\bibfield  {journal} {\bibinfo  {journal} {Nature Communications}\
  }\textbf {\bibinfo {volume} {13}},\ \bibinfo {pages} {4357} (\bibinfo {year}
  {2022})}\BibitemShut {NoStop}%
\bibitem [{\citenamefont {Turton}\ \emph {et~al.}(2018)\citenamefont {Turton},
  \citenamefont {Couchman},\ and\ \citenamefont {Bush}}]{turton2018review}%
  \BibitemOpen
  \bibfield  {author} {\bibinfo {author} {\bibfnamefont {S.}~\bibnamefont
  {Turton}}, \bibinfo {author} {\bibfnamefont {M.}~\bibnamefont {Couchman}}, \
  and\ \bibinfo {author} {\bibfnamefont {J.}~\bibnamefont {Bush}},\ }\href@noop
  {} {\bibfield  {journal} {\bibinfo  {journal} {Chaos}\ }\textbf {\bibinfo
  {volume} {28}},\ \bibinfo {pages} {096111} (\bibinfo {year}
  {2018})}\BibitemShut {NoStop}%
\bibitem [{\citenamefont {Borghesi}\ \emph {et~al.}(2014)\citenamefont
  {Borghesi}, \citenamefont {Moukhtar}, \citenamefont {Labousse}, \citenamefont
  {Eddi}, \citenamefont {Fort},\ and\ \citenamefont {Couder}}]{Borghesi2014}%
  \BibitemOpen
  \bibfield  {author} {\bibinfo {author} {\bibfnamefont {C.}~\bibnamefont
  {Borghesi}}, \bibinfo {author} {\bibfnamefont {J.}~\bibnamefont {Moukhtar}},
  \bibinfo {author} {\bibfnamefont {M.}~\bibnamefont {Labousse}}, \bibinfo
  {author} {\bibfnamefont {A.}~\bibnamefont {Eddi}}, \bibinfo {author}
  {\bibfnamefont {E.}~\bibnamefont {Fort}}, \ and\ \bibinfo {author}
  {\bibfnamefont {Y.}~\bibnamefont {Couder}},\ }\href {\doibase
  10.1103/PhysRevE.90.063017} {\bibfield  {journal} {\bibinfo  {journal}
  {Physical Review E - Statistical, Nonlinear, and Soft Matter Physics}\
  }\textbf {\bibinfo {volume} {90}},\ \bibinfo {pages} {063017} (\bibinfo
  {year} {2014})},\ \Eprint {http://arxiv.org/abs/1412.7701} {arXiv:1412.7701}
  \BibitemShut {NoStop}%
\bibitem [{\citenamefont {Valani}\ and\ \citenamefont
  {Slim}(2018)}]{Valani2018}%
  \BibitemOpen
  \bibfield  {author} {\bibinfo {author} {\bibfnamefont {R.~N.}\ \bibnamefont
  {Valani}}\ and\ \bibinfo {author} {\bibfnamefont {A.~C.}\ \bibnamefont
  {Slim}},\ }\href {\doibase 10.1063/1.5032128} {\bibfield  {journal} {\bibinfo
   {journal} {Chaos}\ }\textbf {\bibinfo {volume} {28}},\ \bibinfo {eid}
  {096114} (\bibinfo {year} {2018})}\BibitemShut {NoStop}%
\bibitem [{\citenamefont {Arbelaiz}\ \emph {et~al.}(2018)\citenamefont
  {Arbelaiz}, \citenamefont {Oza},\ and\ \citenamefont {Bush}}]{Arbelaiz2018}%
  \BibitemOpen
  \bibfield  {author} {\bibinfo {author} {\bibfnamefont {J.}~\bibnamefont
  {Arbelaiz}}, \bibinfo {author} {\bibfnamefont {A.~U.}\ \bibnamefont {Oza}}, \
  and\ \bibinfo {author} {\bibfnamefont {J.~W.~M.}\ \bibnamefont {Bush}},\
  }\href {\doibase 10.1103/PhysRevFluids.3.013604} {\bibfield  {journal}
  {\bibinfo  {journal} {Physical Review Fluids}\ }\textbf {\bibinfo {volume}
  {3}},\ \bibinfo {eid} {013604} (\bibinfo {year} {2018})}\BibitemShut
  {NoStop}%
\bibitem [{\citenamefont {Couchman}\ \emph {et~al.}(2019)\citenamefont
  {Couchman}, \citenamefont {Turton},\ and\ \citenamefont
  {Bush}}]{Couchman2019}%
  \BibitemOpen
  \bibfield  {author} {\bibinfo {author} {\bibfnamefont {M.~M.~P.}\
  \bibnamefont {Couchman}}, \bibinfo {author} {\bibfnamefont {S.~E.}\
  \bibnamefont {Turton}}, \ and\ \bibinfo {author} {\bibfnamefont {J.~W.~M.}\
  \bibnamefont {Bush}},\ }\href {\doibase 10.1017/jfm.2019.293} {\bibfield
  {journal} {\bibinfo  {journal} {Journal of Fluid Mechanics}\ }\textbf
  {\bibinfo {volume} {871}},\ \bibinfo {pages} {212} (\bibinfo {year}
  {2019})}\BibitemShut {NoStop}%
\bibitem [{\citenamefont {Eddi}\ \emph {et~al.}(2012)\citenamefont {Eddi},
  \citenamefont {Moukhtar}, \citenamefont {Perrard}, \citenamefont {Fort},\
  and\ \citenamefont {Couder}}]{Eddi2012}%
  \BibitemOpen
  \bibfield  {author} {\bibinfo {author} {\bibfnamefont {A.}~\bibnamefont
  {Eddi}}, \bibinfo {author} {\bibfnamefont {J.}~\bibnamefont {Moukhtar}},
  \bibinfo {author} {\bibfnamefont {S.}~\bibnamefont {Perrard}}, \bibinfo
  {author} {\bibfnamefont {E.}~\bibnamefont {Fort}}, \ and\ \bibinfo {author}
  {\bibfnamefont {Y.}~\bibnamefont {Couder}},\ }\href@noop {} {\bibfield
  {journal} {\bibinfo  {journal} {Phys. Rev. Lett.}\ }\textbf {\bibinfo
  {volume} {108}} (\bibinfo {year} {2012})}\BibitemShut {NoStop}%
\bibitem [{\citenamefont {Oza}\ \emph {et~al.}(2006)\citenamefont {Oza},
  \citenamefont {Si{\'e}fert}, \citenamefont {Harris}, \citenamefont
  {Mol{\'{a}}{\v{c}}ek},\ and\ \citenamefont {Bush}}]{Oza2019_orbits}%
  \BibitemOpen
  \bibfield  {author} {\bibinfo {author} {\bibfnamefont {A.~U.}\ \bibnamefont
  {Oza}}, \bibinfo {author} {\bibfnamefont {E.}~\bibnamefont {Si{\'e}fert}},
  \bibinfo {author} {\bibfnamefont {D.~M.}\ \bibnamefont {Harris}}, \bibinfo
  {author} {\bibfnamefont {J.}~\bibnamefont {Mol{\'{a}}{\v{c}}ek}}, \ and\
  \bibinfo {author} {\bibfnamefont {J.~W.~M.}\ \bibnamefont {Bush}},\
  }\href@noop {} {\bibfield  {journal} {\bibinfo  {journal} {Physical Review
  Fluids}\ }\textbf {\bibinfo {volume} {2}},\ \bibinfo {pages} {053601}
  (\bibinfo {year} {2006})}\BibitemShut {NoStop}%
\bibitem [{\citenamefont {Papatryfonos}\ \emph
  {et~al.}(2022{\natexlab{a}})\citenamefont {Papatryfonos}, \citenamefont
  {Ruelle}, \citenamefont {Bourdiol}, \citenamefont {Nachbin}, \citenamefont
  {M.},\ and\ \citenamefont {Labousse}}]{Papatryfonos2022}%
  \BibitemOpen
  \bibfield  {author} {\bibinfo {author} {\bibfnamefont {K.}~\bibnamefont
  {Papatryfonos}}, \bibinfo {author} {\bibfnamefont {M.}~\bibnamefont
  {Ruelle}}, \bibinfo {author} {\bibfnamefont {C.}~\bibnamefont {Bourdiol}},
  \bibinfo {author} {\bibfnamefont {A.}~\bibnamefont {Nachbin}}, \bibinfo
  {author} {\bibfnamefont {B.~J.~W.}\ \bibnamefont {M.}}, \ and\ \bibinfo
  {author} {\bibfnamefont {M.}~\bibnamefont {Labousse}},\ }\href@noop {}
  {\bibfield  {journal} {\bibinfo  {journal} {Communication Physics}\ }\textbf
  {\bibinfo {volume} {5}},\ \bibinfo {pages} {142} (\bibinfo {year}
  {2022}{\natexlab{a}})}\BibitemShut {NoStop}%
\bibitem [{\citenamefont {Frumkin}\ \emph {et~al.}(2023)\citenamefont
  {Frumkin}, \citenamefont {M.~Bush},\ and\ \citenamefont
  {Papatryfonos}}]{Papatryfonos2023}%
  \BibitemOpen
  \bibfield  {author} {\bibinfo {author} {\bibfnamefont {V.}~\bibnamefont
  {Frumkin}}, \bibinfo {author} {\bibfnamefont {J.}~\bibnamefont {M.~Bush}}, \
  and\ \bibinfo {author} {\bibfnamefont {K.}~\bibnamefont {Papatryfonos}},\
  }\href@noop {} {\bibfield  {journal} {\bibinfo  {journal} {Phys. Rev. Lett.}\
  } (\bibinfo {year} {2023})}\BibitemShut {NoStop}%
\bibitem [{\citenamefont {Papatryfonos}\ \emph
  {et~al.}(2022{\natexlab{b}})\citenamefont {Papatryfonos}, \citenamefont
  {Vervoort}, \citenamefont {Nachbin},\ and\ \citenamefont
  {Labousse}}]{Bell_Papatryfonos}%
  \BibitemOpen
  \bibfield  {author} {\bibinfo {author} {\bibfnamefont {K.}~\bibnamefont
  {Papatryfonos}}, \bibinfo {author} {\bibfnamefont {L.}~\bibnamefont
  {Vervoort}}, \bibinfo {author} {\bibfnamefont {A.}~\bibnamefont {Nachbin}}, \
  and\ \bibinfo {author} {\bibfnamefont {J.}~\bibnamefont {Labousse},
  \bibfnamefont {M.~Bush}},\ }\href@noop {} {\bibfield  {journal} {\bibinfo
  {journal} {ArXiv preprint}\ } (\bibinfo {year}
  {2022}{\natexlab{b}})}\BibitemShut {NoStop}%
\bibitem [{\citenamefont {Eddi}\ \emph {et~al.}(2009)\citenamefont {Eddi},
  \citenamefont {Decelle}, \citenamefont {Fort},\ and\ \citenamefont
  {Couder}}]{Eddi2009a}%
  \BibitemOpen
  \bibfield  {author} {\bibinfo {author} {\bibfnamefont {A.}~\bibnamefont
  {Eddi}}, \bibinfo {author} {\bibfnamefont {A.}~\bibnamefont {Decelle}},
  \bibinfo {author} {\bibfnamefont {E.}~\bibnamefont {Fort}}, \ and\ \bibinfo
  {author} {\bibfnamefont {Y.}~\bibnamefont {Couder}},\ }\href@noop {}
  {\bibfield  {journal} {\bibinfo  {journal} {Europhys. Lett.}\ }\textbf
  {\bibinfo {volume} {87}} (\bibinfo {year} {2009})}\BibitemShut {NoStop}%
\bibitem [{\citenamefont {Eddi}\ \emph
  {et~al.}(2011{\natexlab{b}})\citenamefont {Eddi}, \citenamefont {Boudaoud},\
  and\ \citenamefont {Couder}}]{Eddi2011b}%
  \BibitemOpen
  \bibfield  {author} {\bibinfo {author} {\bibfnamefont {A.}~\bibnamefont
  {Eddi}}, \bibinfo {author} {\bibfnamefont {A.}~\bibnamefont {Boudaoud}}, \
  and\ \bibinfo {author} {\bibfnamefont {Y.}~\bibnamefont {Couder}},\
  }\href@noop {} {\bibfield  {journal} {\bibinfo  {journal} {Euro. Phys.
  Lett.}\ }\textbf {\bibinfo {volume} {94}} (\bibinfo {year}
  {2011}{\natexlab{b}})}\BibitemShut {NoStop}%
\bibitem [{\citenamefont {Couchman}\ \emph {et~al.}(2022)\citenamefont
  {Couchman}, \citenamefont {Evans},\ and\ \citenamefont
  {Bush}}]{CouchmanEvansBush2022}%
  \BibitemOpen
  \bibfield  {author} {\bibinfo {author} {\bibfnamefont {M.~M.~P.}\
  \bibnamefont {Couchman}}, \bibinfo {author} {\bibfnamefont {D.~J.}\
  \bibnamefont {Evans}}, \ and\ \bibinfo {author} {\bibfnamefont {J.~W.~M.}\
  \bibnamefont {Bush}},\ }\href {\doibase 10.3390/sym14081524} {\bibfield
  {journal} {\bibinfo  {journal} {Symmetry}\ }\textbf {\bibinfo {volume}
  {14}},\ \bibinfo {pages} {1524} (\bibinfo {year} {2022})}\BibitemShut
  {NoStop}%
\bibitem [{\citenamefont {Filoux}\ \emph {et~al.}(2015)\citenamefont {Filoux},
  \citenamefont {Hubert},\ and\ \citenamefont {Vandewalle}}]{Filoux2015}%
  \BibitemOpen
  \bibfield  {author} {\bibinfo {author} {\bibfnamefont {B.}~\bibnamefont
  {Filoux}}, \bibinfo {author} {\bibfnamefont {M.}~\bibnamefont {Hubert}}, \
  and\ \bibinfo {author} {\bibfnamefont {N.}~\bibnamefont {Vandewalle}},\
  }\href {\doibase 10.1103/PhysRevE.92.041004} {\bibfield  {journal} {\bibinfo
  {journal} {Physical Review E}\ }\textbf {\bibinfo {volume} {92}},\ \bibinfo
  {pages} {041004} (\bibinfo {year} {2015})}\BibitemShut {NoStop}%
\bibitem [{\citenamefont {Filoux}\ \emph {et~al.}(2017)\citenamefont {Filoux},
  \citenamefont {Hubert}, \citenamefont {Schlagheck},\ and\ \citenamefont
  {Vandewalle}}]{Filoux2017}%
  \BibitemOpen
  \bibfield  {author} {\bibinfo {author} {\bibfnamefont {B.}~\bibnamefont
  {Filoux}}, \bibinfo {author} {\bibfnamefont {M.}~\bibnamefont {Hubert}},
  \bibinfo {author} {\bibfnamefont {P.}~\bibnamefont {Schlagheck}}, \ and\
  \bibinfo {author} {\bibfnamefont {N.}~\bibnamefont {Vandewalle}},\ }\href
  {\doibase 10.1103/PhysRevFluids.2.013601} {\bibfield  {journal} {\bibinfo
  {journal} {Phys. Rev. Fluids}\ }\textbf {\bibinfo {volume} {2}},\ \bibinfo
  {pages} {013601} (\bibinfo {year} {2017})}\BibitemShut {NoStop}%
\bibitem [{\citenamefont {Vandewalle}\ \emph {et~al.}(2019)\citenamefont
  {Vandewalle}, \citenamefont {Filoux},\ and\ \citenamefont
  {Hubert}}]{Vandewalle_bragg}%
  \BibitemOpen
  \bibfield  {author} {\bibinfo {author} {\bibfnamefont {N.}~\bibnamefont
  {Vandewalle}}, \bibinfo {author} {\bibfnamefont {B.}~\bibnamefont {Filoux}},
  \ and\ \bibinfo {author} {\bibfnamefont {M.}~\bibnamefont {Hubert}},\
  }\href@noop {} {\bibfield  {journal} {\bibinfo  {journal} {arXiv preprint}\ }
  (\bibinfo {year} {2019})}\BibitemShut {NoStop}%
\bibitem [{\citenamefont {Couchman}\ and\ \citenamefont
  {Bush}(2020)}]{Couchman2020}%
  \BibitemOpen
  \bibfield  {author} {\bibinfo {author} {\bibfnamefont {M.~M.~P.}\
  \bibnamefont {Couchman}}\ and\ \bibinfo {author} {\bibfnamefont {J.~W.~M.}\
  \bibnamefont {Bush}},\ }\href@noop {} {\bibfield  {journal} {\bibinfo
  {journal} {Journal of Fluid Mechanics}\ }\textbf {\bibinfo {volume} {903}},\
  \bibinfo {pages} {A49} (\bibinfo {year} {2020})}\BibitemShut {NoStop}%
\bibitem [{\citenamefont {Thomson}\ \emph
  {et~al.}(2020{\natexlab{a}})\citenamefont {Thomson}, \citenamefont
  {Couchman},\ and\ \citenamefont {Bush}}]{Thomson2020}%
  \BibitemOpen
  \bibfield  {author} {\bibinfo {author} {\bibfnamefont {S.~J.}\ \bibnamefont
  {Thomson}}, \bibinfo {author} {\bibfnamefont {M.~M.~P.}\ \bibnamefont
  {Couchman}}, \ and\ \bibinfo {author} {\bibfnamefont {J.~W.~M.}\ \bibnamefont
  {Bush}},\ }\href@noop {} {\bibfield  {journal} {\bibinfo  {journal} {Physical
  Review Fluids}\ }\textbf {\bibinfo {volume} {5}},\ \bibinfo {pages} {083601}
  (\bibinfo {year} {2020}{\natexlab{a}})}\BibitemShut {NoStop}%
\bibitem [{\citenamefont {Thomson}\ \emph
  {et~al.}(2020{\natexlab{b}})\citenamefont {Thomson}, \citenamefont {Durey},\
  and\ \citenamefont {Rosales}}]{Thomson2020a}%
  \BibitemOpen
  \bibfield  {author} {\bibinfo {author} {\bibfnamefont {S.~J.}\ \bibnamefont
  {Thomson}}, \bibinfo {author} {\bibfnamefont {M.}~\bibnamefont {Durey}}, \
  and\ \bibinfo {author} {\bibfnamefont {R.~R.}\ \bibnamefont {Rosales}},\
  }\href {\doibase 10.1098/rspa.2020.0155} {\bibfield  {journal} {\bibinfo
  {journal} {Proceedings of the Royal Society of London Series A}\ }\textbf
  {\bibinfo {volume} {476}},\ \bibinfo {eid} {20200155} (\bibinfo {year}
  {2020}{\natexlab{b}})}\BibitemShut {NoStop}%
\bibitem [{\citenamefont {{Thomson}}\ \emph {et~al.}(2021)\citenamefont
  {{Thomson}}, \citenamefont {{Durey}},\ and\ \citenamefont
  {{Rosales}}}]{Thomson2021}%
  \BibitemOpen
  \bibfield  {author} {\bibinfo {author} {\bibfnamefont {S.~J.}\ \bibnamefont
  {{Thomson}}}, \bibinfo {author} {\bibfnamefont {M.}~\bibnamefont {{Durey}}},
  \ and\ \bibinfo {author} {\bibfnamefont {R.~R.}\ \bibnamefont {{Rosales}}},\
  }\href {\doibase 10.1103/PhysRevE.103.062215} {\bibfield  {journal} {\bibinfo
   {journal} {\pre}\ }\textbf {\bibinfo {volume} {103}},\ \bibinfo {eid}
  {062215} (\bibinfo {year} {2021})}\BibitemShut {NoStop}%
\bibitem [{\citenamefont {Barnes}\ \emph {et~al.}(2020)\citenamefont {Barnes},
  \citenamefont {Pucci},\ and\ \citenamefont {Oza}}]{Barnes2020}%
  \BibitemOpen
  \bibfield  {author} {\bibinfo {author} {\bibfnamefont {L.}~\bibnamefont
  {Barnes}}, \bibinfo {author} {\bibfnamefont {G.}~\bibnamefont {Pucci}}, \
  and\ \bibinfo {author} {\bibfnamefont {A.~U.}\ \bibnamefont {Oza}},\ }\href
  {\doibase 10.5802/crmeca.30} {\bibfield  {journal} {\bibinfo  {journal}
  {{Comptes Rendus M{\'e}canique}}\ }\textbf {\bibinfo {volume} {348}},\
  \bibinfo {pages} {573} (\bibinfo {year} {2020})}\BibitemShut {NoStop}%
\bibitem [{\citenamefont {S{\'a}enz}\ \emph {et~al.}(2021)\citenamefont
  {S{\'a}enz}, \citenamefont {Pucci}, \citenamefont {Turton}, \citenamefont
  {Goujon}, \citenamefont {Rosales}, \citenamefont {Dunkel},\ and\
  \citenamefont {Bush}}]{Saenz2021}%
  \BibitemOpen
  \bibfield  {author} {\bibinfo {author} {\bibfnamefont {P.~J.}\ \bibnamefont
  {S{\'a}enz}}, \bibinfo {author} {\bibfnamefont {G.}~\bibnamefont {Pucci}},
  \bibinfo {author} {\bibfnamefont {S.~E.}\ \bibnamefont {Turton}}, \bibinfo
  {author} {\bibfnamefont {A.}~\bibnamefont {Goujon}}, \bibinfo {author}
  {\bibfnamefont {R.~R.}\ \bibnamefont {Rosales}}, \bibinfo {author}
  {\bibfnamefont {J.}~\bibnamefont {Dunkel}}, \ and\ \bibinfo {author}
  {\bibfnamefont {J.~W.~M.}\ \bibnamefont {Bush}},\ }\href@noop {} {\bibfield
  {journal} {\bibinfo  {journal} {Nature}\ }\textbf {\bibinfo {volume} {596}},\
  \bibinfo {pages} {58} (\bibinfo {year} {2021})}\BibitemShut {NoStop}%
\bibitem [{\citenamefont {Galeano-Rios}\ \emph {et~al.}(2019)\citenamefont
  {Galeano-Rios}, \citenamefont {Milewski},\ and\ \citenamefont
  {Vanden-Broeck}}]{Rios19}%
  \BibitemOpen
  \bibfield  {author} {\bibinfo {author} {\bibfnamefont {C.~A.}\ \bibnamefont
  {Galeano-Rios}}, \bibinfo {author} {\bibfnamefont {P.~A.}\ \bibnamefont
  {Milewski}}, \ and\ \bibinfo {author} {\bibfnamefont {J.-M.}\ \bibnamefont
  {Vanden-Broeck}},\ }\href {\doibase 10.1017/jfm.2019.409} {\bibfield
  {journal} {\bibinfo  {journal} {Journal of Fluid Mechanics}\ }\textbf
  {\bibinfo {volume} {873}},\ \bibinfo {pages} {856} (\bibinfo {year}
  {2019})}\BibitemShut {NoStop}%
\bibitem [{\citenamefont {Rayleigh}(1883)}]{Rayleigh1883}%
  \BibitemOpen
  \bibfield  {author} {\bibinfo {author} {\bibfnamefont {L.}~\bibnamefont
  {Rayleigh}},\ }\href@noop {} {\bibfield  {journal} {\bibinfo  {journal}
  {Phil. Mag.}\ }\textbf {\bibinfo {volume} {16}} (\bibinfo {year}
  {1883})}\BibitemShut {NoStop}%
\bibitem [{\citenamefont {Douady}(1990)}]{Douady1990}%
  \BibitemOpen
  \bibfield  {author} {\bibinfo {author} {\bibfnamefont {S.}~\bibnamefont
  {Douady}},\ }\href@noop {} {\bibfield  {journal} {\bibinfo  {journal} {J.
  Fluid Mech.}\ }\textbf {\bibinfo {volume} {221}},\ \bibinfo {pages} {383}
  (\bibinfo {year} {1990})}\BibitemShut {NoStop}%
\bibitem [{\citenamefont {Kumar}\ and\ \citenamefont
  {Tuckerman}(1994)}]{Kumar1994}%
  \BibitemOpen
  \bibfield  {author} {\bibinfo {author} {\bibfnamefont {K.}~\bibnamefont
  {Kumar}}\ and\ \bibinfo {author} {\bibfnamefont {L.~S.}\ \bibnamefont
  {Tuckerman}},\ }\href@noop {} {\bibfield  {journal} {\bibinfo  {journal}
  {Journal of Fluid Mechanics}\ }\textbf {\bibinfo {volume} {279}},\ \bibinfo
  {pages} {49} (\bibinfo {year} {1994})}\BibitemShut {NoStop}%
\bibitem [{\citenamefont {Kumar}(1996)}]{Kumar1996}%
  \BibitemOpen
  \bibfield  {author} {\bibinfo {author} {\bibfnamefont {K.}~\bibnamefont
  {Kumar}},\ }\href@noop {} {\bibfield  {journal} {\bibinfo  {journal}
  {Proceedings of the Royal Society A}\ }\textbf {\bibinfo {volume} {452}},\
  \bibinfo {pages} {1113} (\bibinfo {year} {1996})}\BibitemShut {NoStop}%
\bibitem [{\citenamefont {Proti{\`{e}}re}\ \emph {et~al.}(2006)\citenamefont
  {Proti{\`{e}}re}, \citenamefont {Boudaoud},\ and\ \citenamefont
  {Couder}}]{Protiere2006}%
  \BibitemOpen
  \bibfield  {author} {\bibinfo {author} {\bibfnamefont {S.}~\bibnamefont
  {Proti{\`{e}}re}}, \bibinfo {author} {\bibfnamefont {A.}~\bibnamefont
  {Boudaoud}}, \ and\ \bibinfo {author} {\bibfnamefont {Y.}~\bibnamefont
  {Couder}},\ }\href {\doibase 10.1017/S0022112006009190} {\bibfield  {journal}
  {\bibinfo  {journal} {Journal of Fluid Mechanics}\ }\textbf {\bibinfo
  {volume} {554}},\ \bibinfo {pages} {85} (\bibinfo {year} {2006})}\BibitemShut
  {NoStop}%
\bibitem [{\citenamefont {Tadrist}\ \emph {et~al.}(2018)\citenamefont
  {Tadrist}, \citenamefont {Shim}, \citenamefont {Gilet},\ and\ \citenamefont
  {Schlagheck}}]{tadrist2018faraday}%
  \BibitemOpen
  \bibfield  {author} {\bibinfo {author} {\bibfnamefont {L.}~\bibnamefont
  {Tadrist}}, \bibinfo {author} {\bibfnamefont {J.-B.}\ \bibnamefont {Shim}},
  \bibinfo {author} {\bibfnamefont {T.}~\bibnamefont {Gilet}}, \ and\ \bibinfo
  {author} {\bibfnamefont {P.}~\bibnamefont {Schlagheck}},\ }\href@noop {}
  {\bibfield  {journal} {\bibinfo  {journal} {Journal of Fluid Mechanics}\
  }\textbf {\bibinfo {volume} {848}},\ \bibinfo {pages} {906} (\bibinfo {year}
  {2018})}\BibitemShut {NoStop}%
\bibitem [{\citenamefont {Labousse}(2014)}]{Labousse_Thesis}%
  \BibitemOpen
  \bibfield  {author} {\bibinfo {author} {\bibfnamefont {M.}~\bibnamefont
  {Labousse}},\ }\emph {\bibinfo {title} {Etude d'une dynamique à m\'emoire de
  chemin: une exp\'erimentation th\'eorique}},\ \href@noop {} {Ph.D. thesis},\
  \bibinfo  {school} {Universit\'e Pierre et Marie Curie-Paris VI} (\bibinfo
  {year} {2014})\BibitemShut {NoStop}%
\bibitem [{\citenamefont {Tambasco}\ and\ \citenamefont
  {Bush}(2018)}]{Tambasco2018}%
  \BibitemOpen
  \bibfield  {author} {\bibinfo {author} {\bibfnamefont {L.}~\bibnamefont
  {Tambasco}}\ and\ \bibinfo {author} {\bibfnamefont {J.}~\bibnamefont
  {Bush}},\ }\href@noop {} {\bibfield  {journal} {\bibinfo  {journal} {Chaos
  (Focus Issue: Hydrodynamic Quantum Analogs)}\ }\textbf {\bibinfo {volume}
  {28}},\ \bibinfo {pages} {096115} (\bibinfo {year} {2018})}\BibitemShut
  {NoStop}%
\bibitem [{\citenamefont {Olver}\ \emph {et~al.}(2010)\citenamefont {Olver},
  \citenamefont {Lozier}, \citenamefont {R.F.},\ and\ \citenamefont
  {Clark}}]{NIST}%
  \BibitemOpen
  \bibfield  {author} {\bibinfo {author} {\bibfnamefont {F.}~\bibnamefont
  {Olver}}, \bibinfo {author} {\bibfnamefont {D.}~\bibnamefont {Lozier}},
  \bibinfo {author} {\bibfnamefont {B.}~\bibnamefont {R.F.}}, \ and\ \bibinfo
  {author} {\bibfnamefont {C.}~\bibnamefont {Clark}},\ }\href@noop {} {\emph
  {\bibinfo {title} {NIST Handbook of Mathematical Functions}}}\ (\bibinfo
  {publisher} {Cambridge University Press},\ \bibinfo {year} {2010})\
  Chap.~\bibinfo {chapter} {10}, p.\ \bibinfo {pages} {247}\BibitemShut
  {NoStop}%
\end{thebibliography}%
\end{document}